\documentclass[prd,preprintnumbers,twocolumn,floatfix]{revtex4}
\usepackage{amsmath,epsfig}

\newcommand{\rf}[4]{{#1} {\bf #2}, #3 (#4)}

\newcommand{\pr}{Phys.\ Rev.}
\newcommand{\physrl}{Phys.\ Rev.\ Lett.}
\newcommand{\physl}{Phys.\ Lett.}

\newcommand{\np}{Nucl.\ Phys.}
\newcommand{\npps}[3]{Nucl.\ Phys. {\bf B} (Proc.\ Suppl.) {\bf #1}, 
	#2 (#3)}

\newcommand{\etal}{\emph{et al.}}

\newcommand{\del}{\partial}
\newcommand{\lapI}{$\partial^2$(I) }
\newcommand{\lapII}{$\partial^2$(II) }
\newcommand{\lapIII}{$\partial^2$(III) }

\newcommand{\adj}{\dagger}
\newcommand{\tr}{\text{Tr}}

\newcommand{\gambar}{\overline{\gamma}}
\newcommand{\chibar}{\overline{\chi}}
\newcommand{\deltabar}{\overline{\delta}}

\newcommand{\oa}[1]{\ensuremath{{\cal O}(a^{#1})}}
\newcommand{\oag}[2]{\ensuremath{{\cal O}(a^{#1}g^{#2})}}

\newcommand{\eref}[1]{Eq.~(\ref{#1})}

\begin{document}

\title{Lattice quark propagator with staggered quarks in Landau and 
	Laplacian gauges}
 
\date{\today}

\author{Patrick O.~Bowman}
\author{Urs M.~Heller}
\affiliation{Department of Physics and 
School for Computational Science and
Information Technology, Florida State University, Tallahassee FL 32306-4120, USA}
\author{Anthony G.~Williams}
\affiliation{Special Research Centre for the Subatomic Structure of Matter 
and \\ The Department of Physics and Mathematical Physics,
Adelaide University, Australia 5005}

\begin{abstract}
We report on the lattice quark propagator using standard and improved 
Staggered quark actions, with the standard, Wilson gauge action.  The standard 
Kogut-Susskind action has errors of \oa{2} while the ``Asqtad'' action has 
\oa{4}, \oag{2}{2} errors.  The quark propagator is interesting for studying
the phenomenon of dynamical chiral symmetry breaking and as a test-bed for
improvement.  
Gauge dependent quantities from lattice simulations may be affected by Gribov
copies.  We explore this by studying the quark propagator in both Landau and 
Laplacian gauges.  Landau and Laplacian gauges are found to produce very 
similar results for the quark propagator.
\end{abstract}

\preprint{FSU-CSIT-02-10}
\preprint{ADP-02-67/T507}

\maketitle

\section{Introduction}

The quark propagator lies at the heart of most QCD physics.  In the low 
momentum region
it exhibits dynamical chiral symmetry breaking (which cannot be seen from
perturbation theory) and at high momentum can be used to extract the 
running quark mass~\cite{Aok99,Bec00b} (which cannot be extracted directly from
experiment).  In lattice QCD, quark propagators are tied together to extract
hadron masses.  Lattice gauge theory provides a way to study the quark
propagator nonperturbatively, possibly as a way of calculating the chiral
condensate and $\Lambda_{\text{QCD}}$, and in turn, such a study can provide 
technical insight into lattice gauge theory.

We study the quark propagator using the Kogut-Susskind (KS) fermion
action, which has \oa{2} errors, and an improved staggered action, 
Asqtad~\cite{Org99}, which has errors of \oa{4}, \oag{2}{2}.  These choices
complement other studies using Clover~\cite{Sku01a, Sku01b} and 
Overlap~\cite{overlap} quarks.  
We are required to gauge fix and we choose the ever popular Landau gauge and 
the interesting Laplacian gauge~\cite{Vin92,Vin95}.
Laplacian gauge fixing is an unambiguous gauge fixing and, although it is 
difficult to understand perturbatively, it is equivalent to Landau gauge 
in the asymptotic region.  It has been used to study the gluon 
propagator~\cite{Ale01,Ale02,lapglu}.

In $SU(N)$ there are various ways to implement a Laplacian gauge fixing.  
Three varieties of Laplacian gauge fixing are used, and these form three 
different, but related gauges.  This is briefly discussed in 
section~\ref{sec:gauge}.  For a more detailed discussion, see 
Ref.~\cite{lapglu}.

The quark propagator was calculated on 80, $16^3 \times 32$ configurations 
generated with the standard Wilson gluon action at $\beta = 5.85$ 
($a = 0.130$ fm)\footnote{Some of numbers in this report differ
slightly from those published elsewhere~\cite{proc_quark01} due to a revised
value for the lattice spacing.}.
We have used
six quark masses: $am = 0.075$, 0.0625, 0.05, 0.0375, 0.025 and 0.0125 
(114 to 19 MeV).

\section{Lattice Quark Propagator}

In the continuum, Lorentz invariance allows us to decompose the full propagator
into Dirac vector and scalar pieces
\begin{equation}
S^{-1}(p^2) = i A(p^2) \gamma \cdot p + B(p^2)
\end{equation}
or, alternatively,
\begin{equation}
S^{-1}(p^2) = Z^{-1}(p^2) [i \gamma \cdot p + M(p^2)].
\end{equation}
This is the bare propagator which, once regularized, is related to the 
renormalized propagator through the renormalization constant
\begin{equation}
S(a;p^2) = Z_2(a;\mu) S^{\text{ren}}(\mu;p^2),
\end{equation}
where $a$ is some regularization parameter, e.g.\ lattice spacing.
Asymptotic freedom implies that, as $p^2 \rightarrow \infty$, $S(p^2)$ reduces 
to the free propagator
\begin{equation}
\label{eq:free_quark}
S^{-1}(p^2) \rightarrow  i\gamma \cdot p + m_0,
\end{equation}
where $m_0$ is the bare quark mass.

The tadpole improved, tree-level form of the KS quark propagator is
\begin{equation}
\label{eq:free_ks}
S_{\alpha\beta}^{-1}(p;m) = u_0 i \sum_\mu {(\gambar_\mu)}_{\alpha\beta} 
   \sin(p_\mu) + m\deltabar_{\alpha\beta}
\end{equation}
where $p_\mu$ is the discrete lattice momentum given by 
\begin{equation}
p_\mu = \frac{2\pi n_\mu}{aL_\mu} \qquad n_\mu \in 
	\Bigl[ \frac{-L_\mu}{4}, \frac{L_\mu}{4} \Bigr).
\end{equation}
For the tadpole factor, we employ the plaquette measure,
\begin{equation}
u_0 = \Bigl( \frac{1}{3} \text{Re} \tr \langle P_{\mu\nu} \rangle 
    \Bigr)^{\frac{1}{4}}.
\end{equation}
We give a detailed discussion of the notation used for the staggered quark
actions in an appendix.
As a convienient short-hand we define a new momentum variable for the KS quark
propagator,
\begin{equation}
\label{eq:mom_KS}
q_\mu \equiv \sin(p_\mu).
\end{equation}
We can then decompose the inverse propagator\footnote{This is intended to be 
merely illustrative; in practice we never actually invert the propagator.
See appendix for details.}
\begin{gather}
\label{eq:extract_Z}
Z^{-1}(q) = \frac{1}{16N_c iq^2} \tr\{\gambar\cdot q S^{-1}\} \\
\label{eq:extract_M}
M(q) = \frac{1}{16N_c} \tr\{S^{-1} \},
\end{gather}
where the factor of 16 comes from the trace over the spin-flavor indices of 
the staggered quarks and $N_c$ from the trace over color.  Comparing 
Eqs.~(\ref{eq:free_quark}) and~(\ref{eq:free_ks}) we see that
dividing out $q^2$ in \eref{eq:extract_Z} is analagous to dividing out
$p^2$ in the continuum and ensures that that $Z$ has the correct asymptotic
behavior.  So by considering the propagator as a function of $q_\mu$, we 
ensure that the lattice quark propagator has the correct tree-level form, 
i.e.,
\begin{equation}
S^\text{tree}(q_\mu) = \frac{1}{i\gambar\cdot q + m}
\end{equation}
and hopefully better approximates its continuum behavior.  This is the same
philosophy that has been used in studies of the gluon propagator~\cite{gluon}
and \eref{eq:mom_KS} was used to define the momentum in Ref.~\cite{Bec00b}.

The Asqtad quark action~\cite{Org99} is a fat-link Staggered action using
three-link, five-link and seven-link staples to minimize flavor changing 
interactions along with the three-link Naik
term~\cite{Nai89} (to correct the dispersion relation) and planar five-link 
Lepage term~\cite{Lep99} (to correct the IR).  The 
coefficients are tadpole improved and tuned to remove all tree-level \oa{2} 
errors.  This action was motivated by the desire to improve flavor symmetry, 
but has also been reported to have good rotational properties.
  
The quark propagator with this action has the tree-level form
\begin{equation}
S_{\alpha\beta}^{-1}(p;m) = u_0 i \sum_\mu {(\gambar_\mu)}_{\alpha\beta}
   \sin(p_\mu) \bigl[ 1 + \frac{1}{6} \sin^2(p_\mu) \bigr] 
   + m\deltabar_{\alpha\beta},
\end{equation}
so we repeat the above analysis, this time defining
\begin{equation}
\label{eq:mom_Asqtad}
q_\mu \equiv \sin(p_\mu) \bigl[ 1 + \frac{1}{6} \sin^2(p_\mu) \bigr].
\end{equation}
Finally, it should be noted that both actions get contributions from tadpoles,
which can be seen in the tree-level behaviors of the two invariants
\begin{align}
Z^\text{tree} &= \frac{1}{u_0} \\
M^\text{tree} &= \frac{m_0}{u_0}
\end{align}
so inserting the tadpole factors provides the correct normalization.

\section{Gauge Fixing}
\label{sec:gauge}

We consider the quark propagator in Landau and Laplacian gauges.  Landau gauge
fixing is performed by enforcing the Lorentz gauge condition, 
$\sum_\mu \del_\mu A_\mu(x) = 0$ on a configuration by configuration basis.
This is achieved by maximizing the functional,
\begin{equation}
{\cal F} = \frac{1}{2} \sum_{x,\mu} \tr \bigl\{ U_\mu(x) 
	+ U_\mu^\adj(x) \bigr\},
\end{equation}
by, in this case, a Fourier accelerated, steepest-descents 
algorithm~\cite{Dav88}.  There are, in general, many such
maxima and these are called lattice Gribov copies.  While this ambiguity has
produced no identified artefacts in QCD, in 
principle it remains a source of uncontrolled systematic error.

Laplacian gauge is a nonlinear gauge fixing that respects rotational
invariance, has been seen to be smooth, yet is free of Gribov ambiguity.
It is also computationally cheaper then Landau gauge.  There is, however, more
than one way of obtaining such a gauge fixing in $SU(N)$. 
The three implementations of Laplacian gauge fixing employed here are
(in our notation):
\begin{enumerate}
\item \lapI gauge (QR decomposition), used by Alexandrou 
	\etal~\cite{Ale01}.
\item \lapII gauge, where the Laplacian gauge transformation is projected 
	onto $SU(3)$ by maximising its trace~\cite{lapglu}.  
\item \lapIII gauge (Polar decomposition), the original prescription described
 	in Ref.~\cite{Vin92} and tested in Ref.~\cite{Vin95}.
\end{enumerate}
All three versions reduce to the same gauge in SU(2).  For a more detailed 
discussion, see Ref.~\cite{lapglu}.

\section{Analysis of Lattice Artefacts}

\subsection{Tree-level Correction}

As mentioned above, the idea of ``kinematic'' or ``tree-level'' correction 
has been used widely
in studies of the gluon propagator~\cite{gluon} and the 
quark propagator~\cite{Sku01a,Sku01b,overlap} and we investigate its
application to our quark propagators.  For the moment we shall restrict 
ourselves to Landau gauge.
To help us understand the lattice artefacts, we separate the data into momenta
lying entirely on a spatial cartesian direction (squares), along the temporal 
direction (triangles), the four-diagonal (diamonds) or some other combination
of directions (circles).

\begin{figure}[h]
\begin{center}
\epsfig{figure=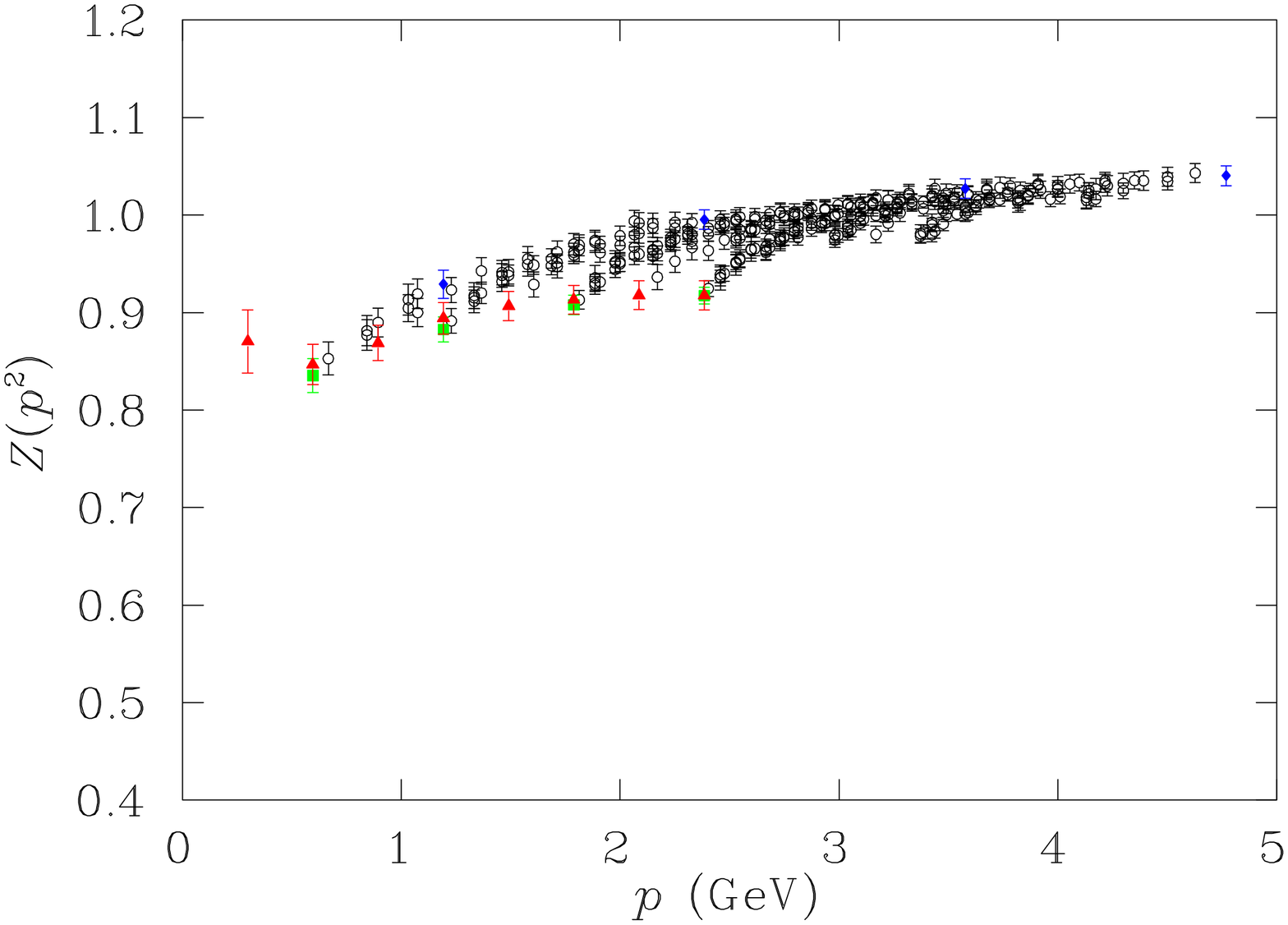,angle=90,width=8cm}
\end{center}
\begin{center}
\epsfig{figure=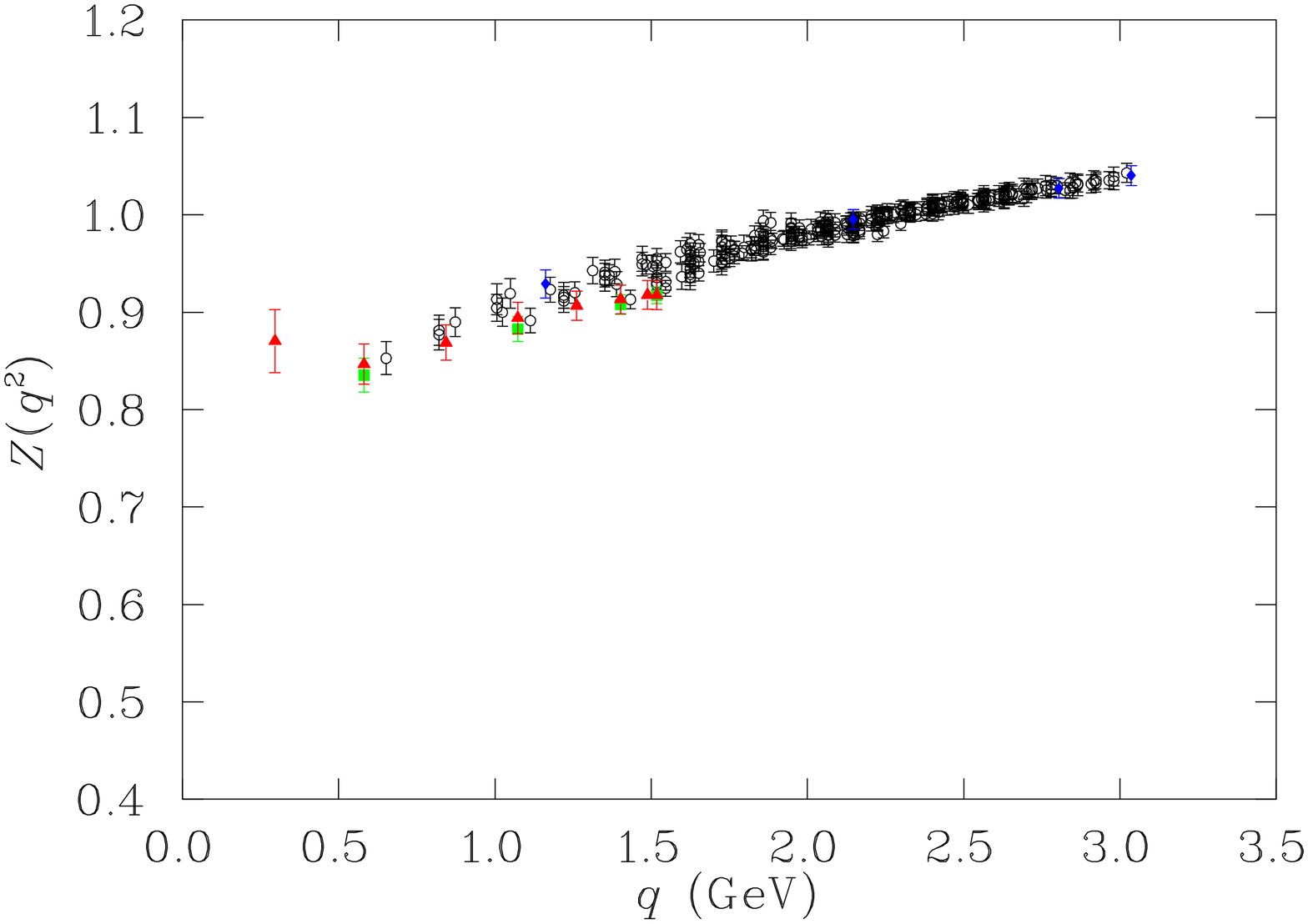,angle=90,width=8cm}
\end{center}
\caption{$Z$ function for quark mass $ma = 0.05$ ($m \simeq 76$ MeV) for the KS
action in Landau gauge.  Top figure is plotted using the standard lattice 
momentum, $p$ and the bottom uses the ``action'' momentum, $q$.  Note that 
this choice affects only the horizontal scale.}
\label{fig:lan_ks_Zm05_comp}
\end{figure}

\begin{figure}[h]
\begin{center}
\epsfig{figure=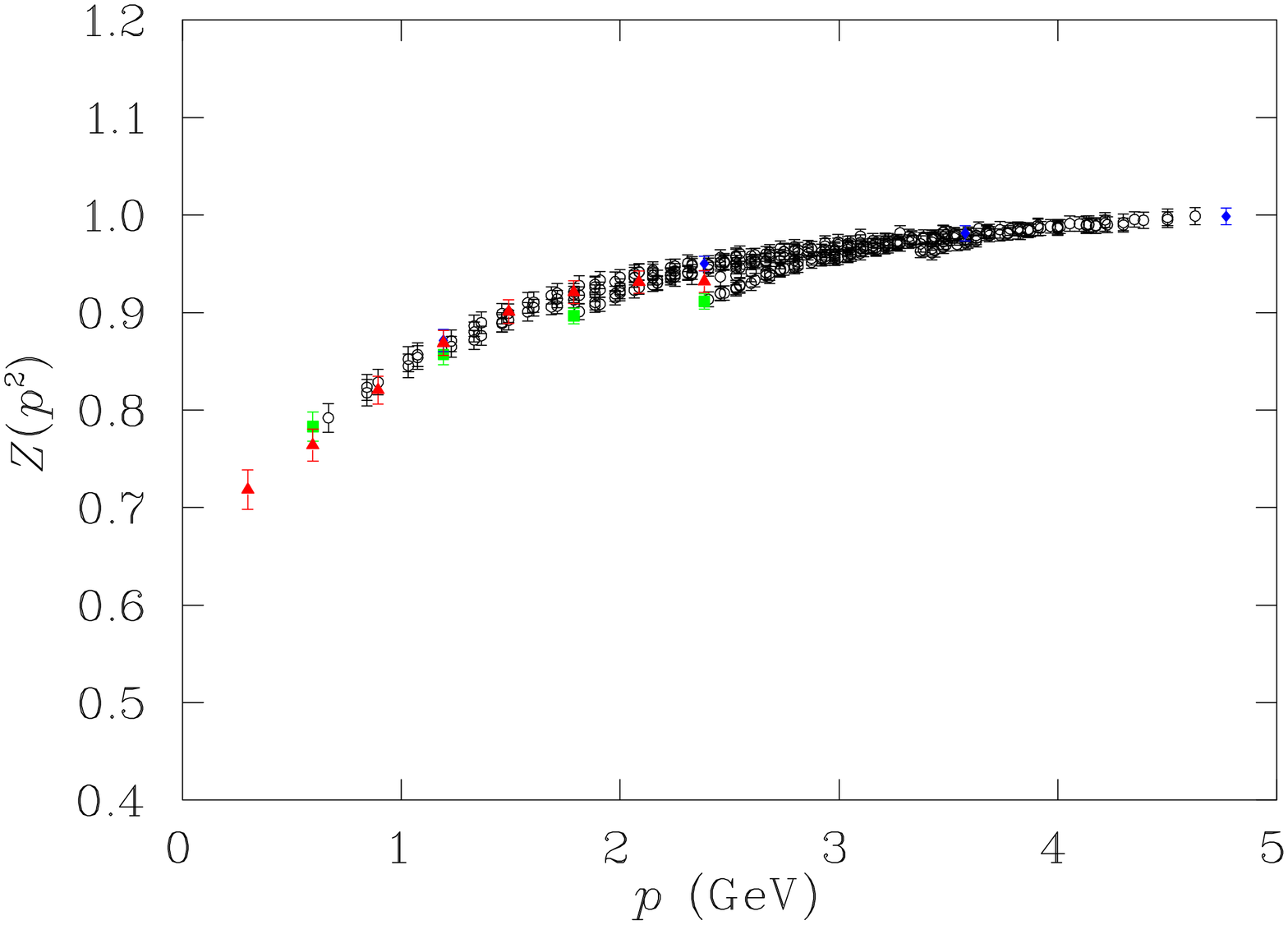,angle=90,width=8cm}
\end{center}
\begin{center}
\epsfig{figure=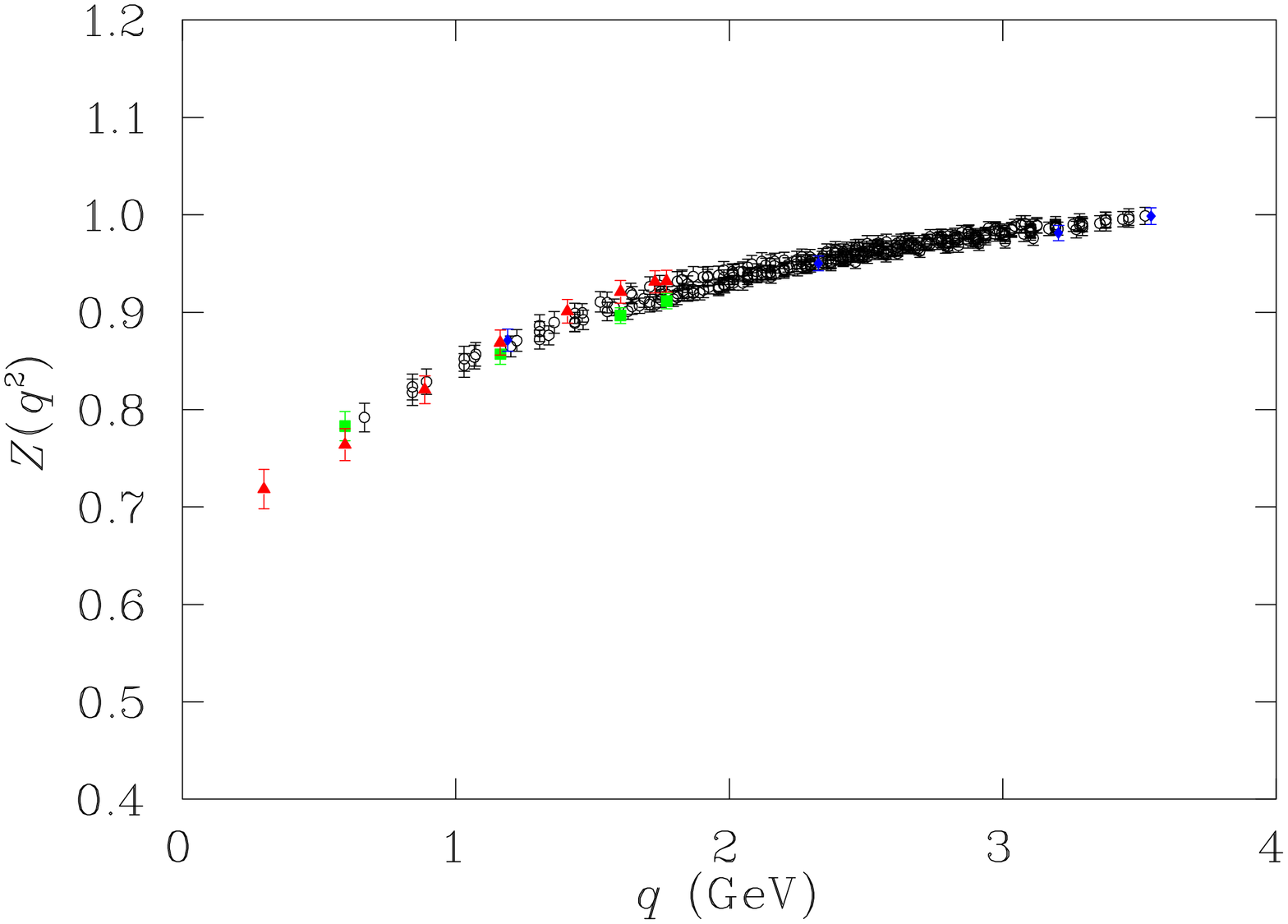,angle=90,width=8cm}
\end{center}
\caption{$Z$ function for quark mass $ma = 0.05$ ($m \simeq 76$ MeV) for the
Asqtad action in Landau gauge.  Top figure is plotted using the standard 
lattice momentum, $p$ and the bottom uses the ``action'' momentum, $q$.}
\label{fig:lan_asq_Zm05_comp}
\end{figure}

The $Z$ function is plotted for the KS action in 
Fig.~\ref{fig:lan_ks_Zm05_comp}, comparing the results using
$p$ and $q$.  In the top of Fig.~\ref{fig:lan_ks_Zm05_comp} we see 
substantial hypercubic artefacts (in particular look at the difference 
between the diamond and the triangle at around 2.5 GeV).  We can suggest
that this is caused by violation of rotational symmetry because the agreement
between triangles and squares suggests that finite volume effects are small
in the region of interest.  In the plot below, where $q$ has been used, we
see some restoration of rotational symmetry.

The same study is made for the Asqtad action in 
Fig.~\ref{fig:lan_asq_Zm05_comp}.  In both cases this action shows a 
substantial improvement over the KS action, and when we plot using $q$, the 
momentum defined by the action, rotational asymmetry is reduced to the level
of the statistical errors.

\begin{figure}[h]
\begin{center}
\epsfig{figure=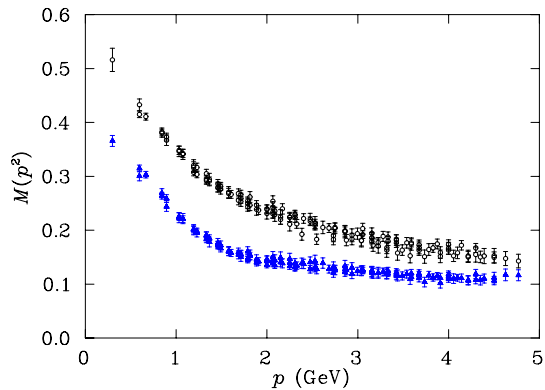,angle=90,width=8cm}
\end{center}
\begin{center}
\epsfig{figure=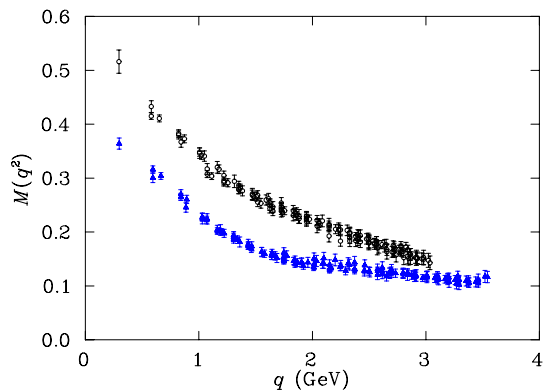,angle=90,width=8cm}
\end{center}
\caption{The quark mass function for quark mass $ma = 0.05$ 
($m \simeq 76$ MeV) for the KS action (open circles) and the
Asqtad action (solid triangles) in Landau gauge.  Top figure is  plotted using
the standard lattice momentum, and the bottom using
the ``action'' momentum.}
\label{fig:lan_comp_m05_comp}
\end{figure}

It is less clear which momentum variable should be used for the mass function,
so for consistency we use $q$, as for the $Z$ function.  The effect of this is
shown in Fig.~\ref{fig:lan_comp_m05_comp}.  For ease of comparison, both sets
of data have been cylinder cut~\cite{gluon}.  In the case of the mass function,
the choice of momentum will actually make little difference to our results.

\subsection{Comparison of the actions}

\begin{figure}[h]
\begin{center}
\epsfig{figure=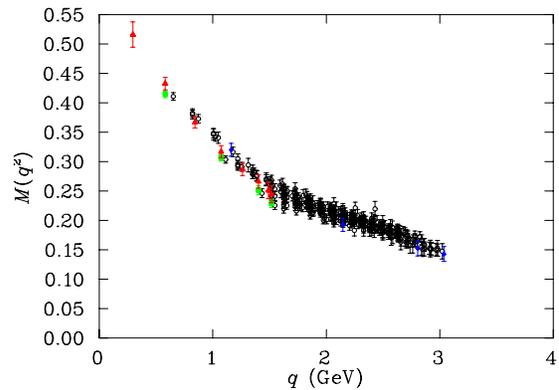,angle=90,width=8cm}
\end{center}
\begin{center}
\epsfig{figure=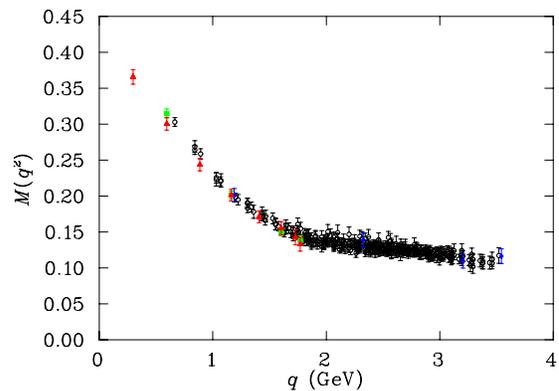,angle=90,width=8cm}
\end{center}
\caption{Mass function for quark mass $ma = 0.05$ ($m \simeq 76$ MeV), KS
action (top) and Asqtad action (bottom) in Landau gauge.}
\label{fig:lan_comp_m05all}
\end{figure}

In Fig.~\ref{fig:lan_comp_m05all} the mass function is plotted, in Landau
gauge, for both actions with quark mass $ma = 0.05$.  This time there have
been no data cuts.  We see that the KS action gives a much larger value for 
M(0) than the Asqtad action and is slower to approach asymptotic behavior.  
Asqtad also shows slightly better rotational symmetry.

\begin{figure}[h]
\begin{center}
\epsfig{figure=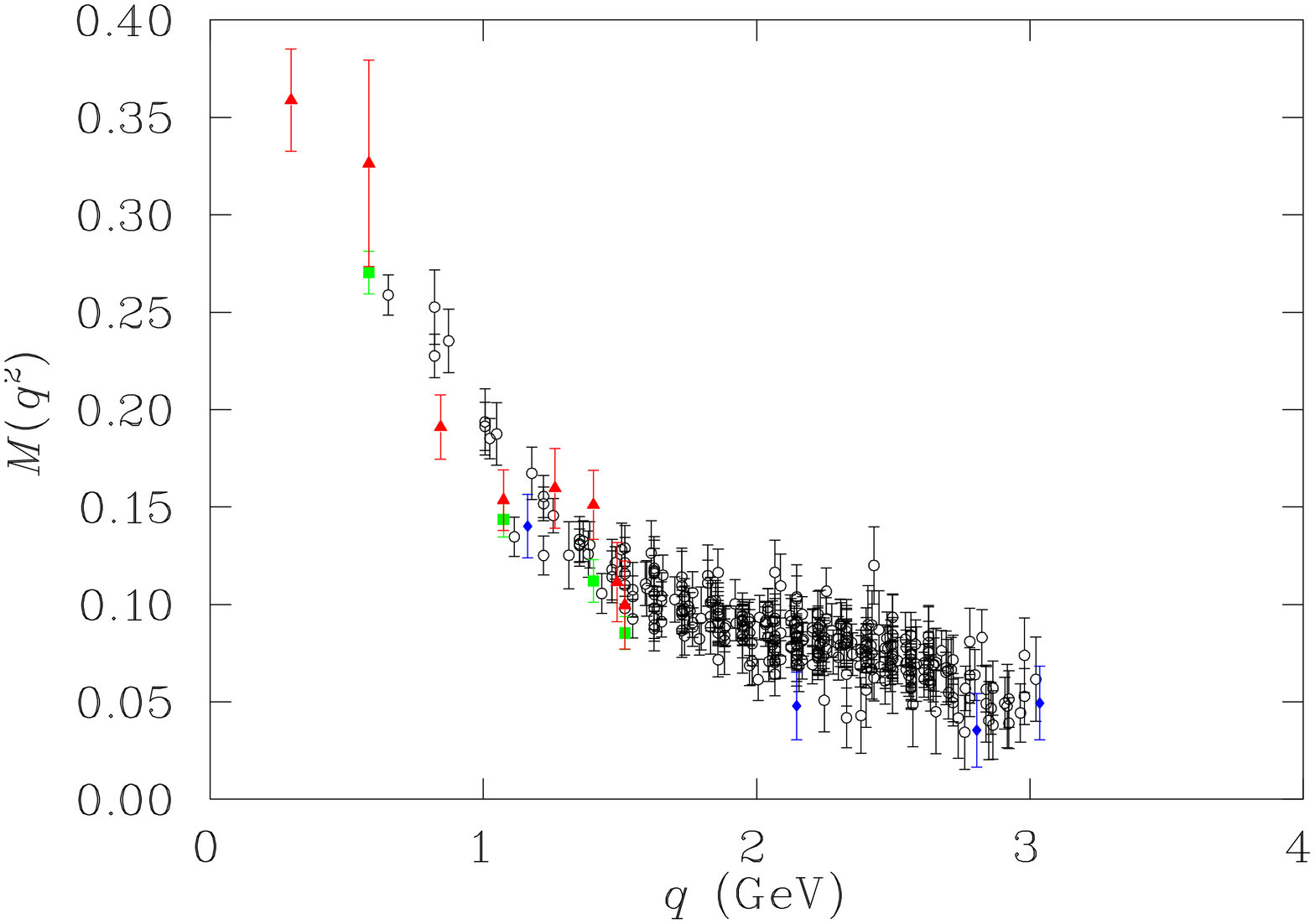,angle=90,width=8cm}
\end{center}
\begin{center}
\epsfig{figure=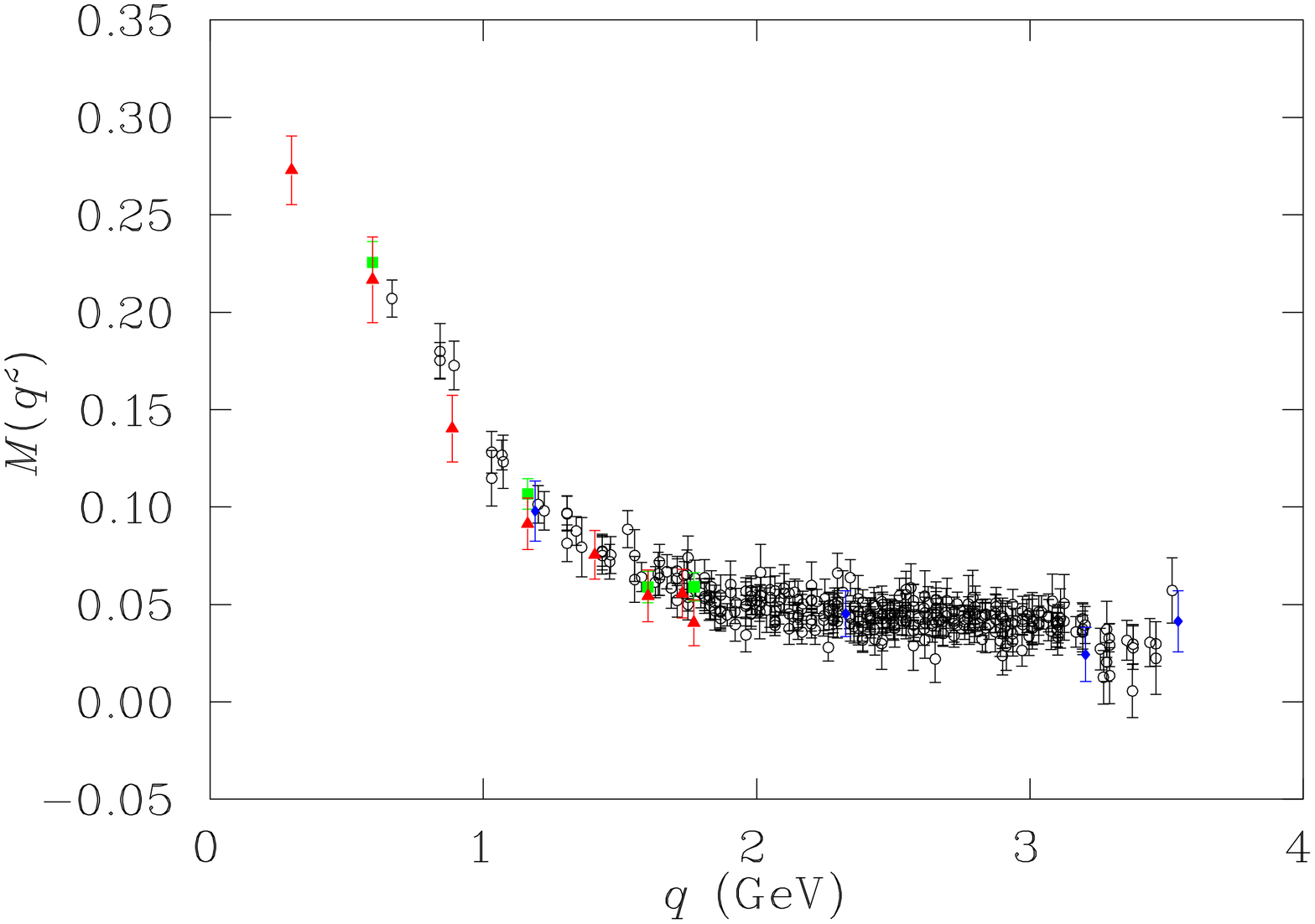,angle=90,width=8cm}
\end{center}
\caption{Mass function for quark mass $ma = 0.0125$ ($m \simeq 19$ MeV), KS
action (top) and Asqtad action (bottom) in Landau gauge.  In both cases, 
lowering the quark mass increases the amount of noise, but the Asqtad action
seems to be affected less than the KS action.  Note that the y-axis for the
bottom figure starts below zero.}
\label{fig:lan_comp_m0125all}
\end{figure}

Looking back at Figs.~\ref{fig:lan_ks_Zm05_comp} 
and~\ref{fig:lan_asq_Zm05_comp} we see that the Asqtad action displays clearly
better rotational symmetry in the quark $Z$ function and, curiously, improved
infrared behavior as well.  The Asqtad action also displays a better approach 
to asymptotic behavior, approaching one in the ultraviolet.  The relative 
improvement increases as the quark mass decreases.  
In Fig.~\ref{fig:lan_comp_m0125all} we compare the mass function for the two
actions at $ma = 0.0125$, the lowest mass studied here.  The low quark mass 
has introduced less noise into the propagator with the Asqtad action than 
with the KS action.

\subsection{Comparitive performance of Landau and Laplacian gauges}

\begin{figure}[h]
\begin{center}
\epsfig{figure=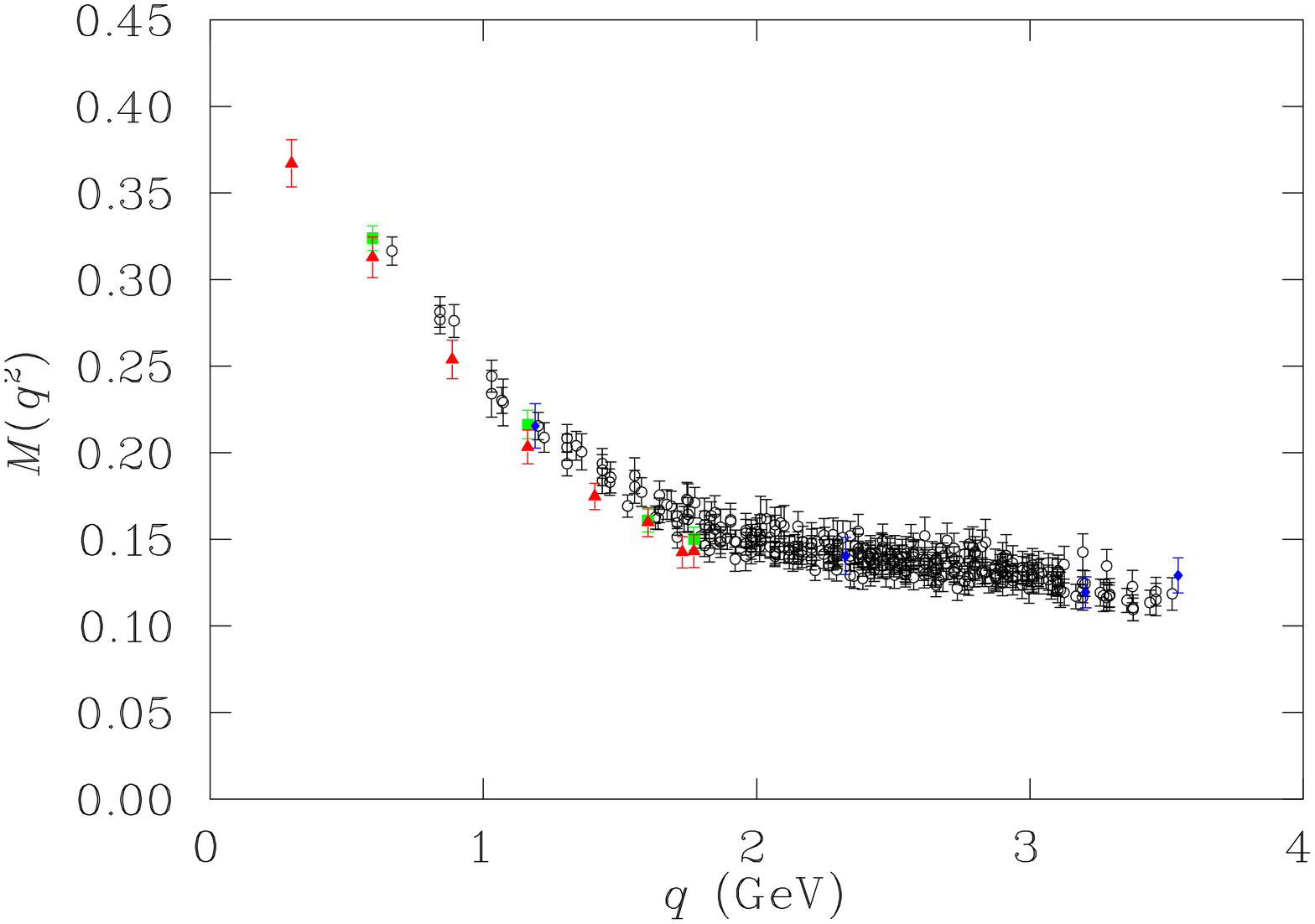,angle=90,width=8cm}
\end{center}
\begin{center}
\epsfig{figure=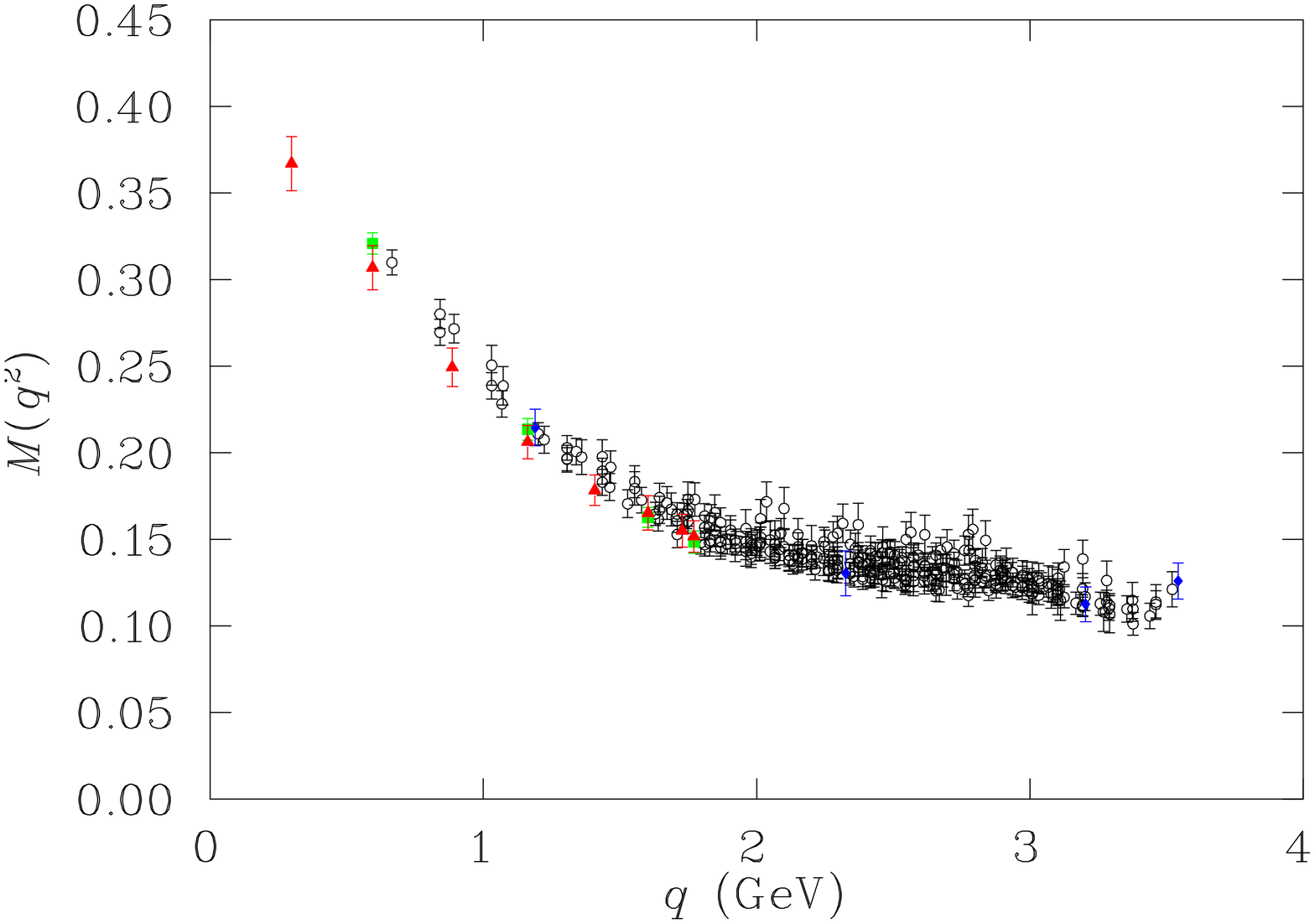,angle=90,width=8cm}
\end{center}
\caption{Mass function for quark mass $ma = 0.05$ ($m \simeq 76$ MeV), Asqtad
action in \lapI and \lapII gauges.  Comparing with 
Fig.~\ref{fig:lan_comp_m05all} we see that Landau, \lapI and \lapII gauges 
yield similar results.}
\label{fig:comp_asq_m05all}
\end{figure}

Fig.~\ref{fig:comp_asq_m05all} shows the mass function for the Asqtad action
in \lapI and \lapII gauges and it should be compared with the equivalent
Landau gauge result in Fig.~\ref{fig:lan_comp_m05all}.  We see firstly
that these three gauges give very similar results (we shall investigate this 
in more detail later) and secondly that they give similar performance in
terms of rotational symmetry and statistical noise.  Looking more closely, we
can see that the Landau gauge gives a slightly cleaner signal at this lattice 
spacing.

\begin{figure}[h]
\begin{center}
\epsfig{figure=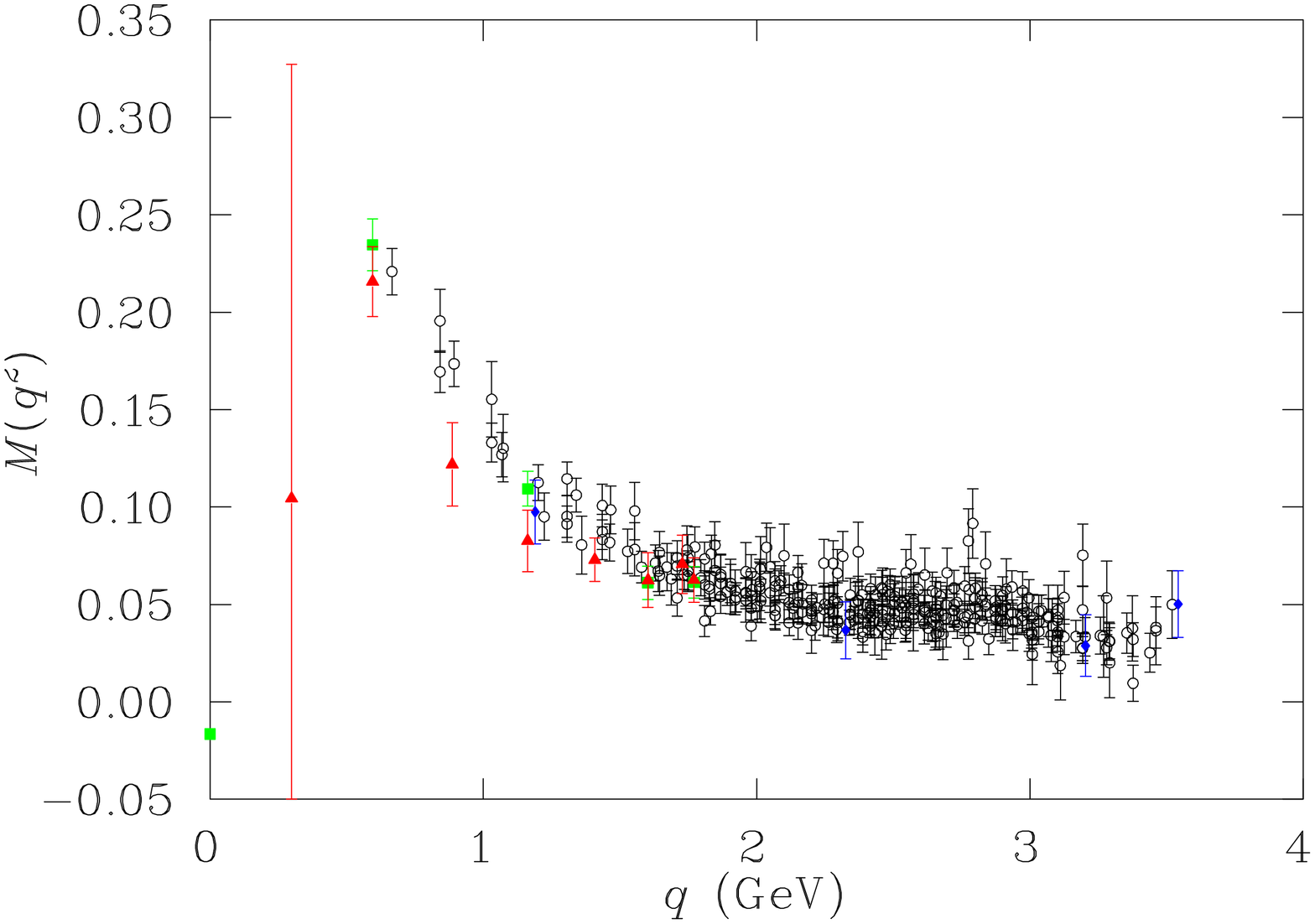,angle=90,width=8cm}
\end{center}
\caption{Mass function for quark mass $ma = 0.0125$ ($m \simeq 19$ MeV), Asqtad
action in \lapII gauge.  We see a lot of infrared noise at this low quark
mass in this gauge.  Compare with the Landau gauge result in 
Fig.~\ref{fig:lan_comp_m0125all}.}
\label{fig:lap2_asq_m0125all}
\end{figure}

Landau gauge seems to respond somewhat better than \lapII gauge to vanishing
quark mass; compare Fig.~\ref{fig:lap2_asq_m0125all} with 
Fig.~\ref{fig:lan_comp_m0125all}.  In Fig.~\ref{fig:lap2_asq_m0125all} we see
large errors in the infrared region and points along the temporal axis lying
below the bulk of the data.  These are indicators of finite volume effects, an
unexpected result given that earlier gluon propagator 
studies~\cite{Ale01,Ale02} appear to conclude that 
Laplacian gauge is less sensitive to volume than Landau 
gauge.

\begin{figure}[h]
\begin{center}
\epsfig{figure=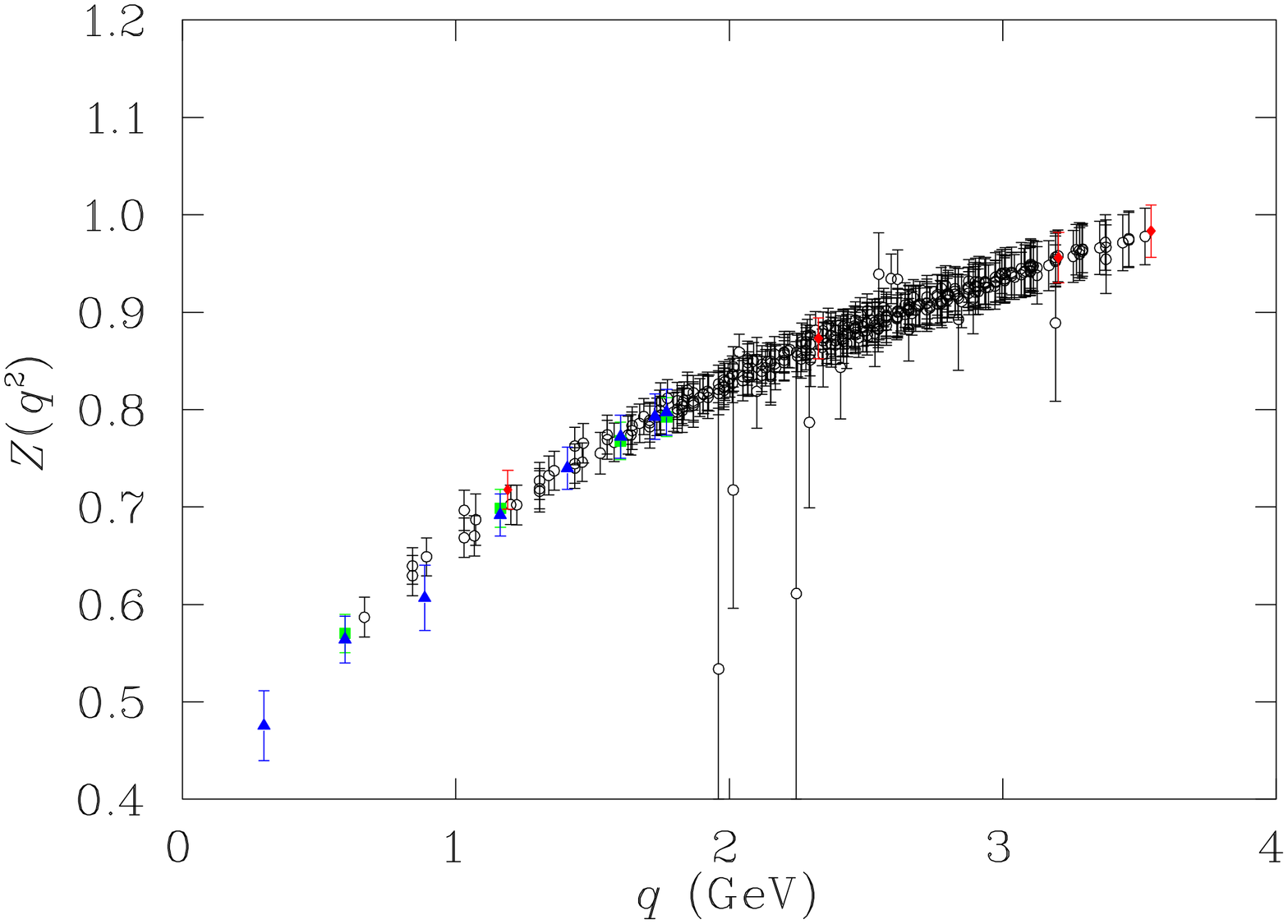,angle=90,width=8cm}
\end{center}
\begin{center}
\epsfig{figure=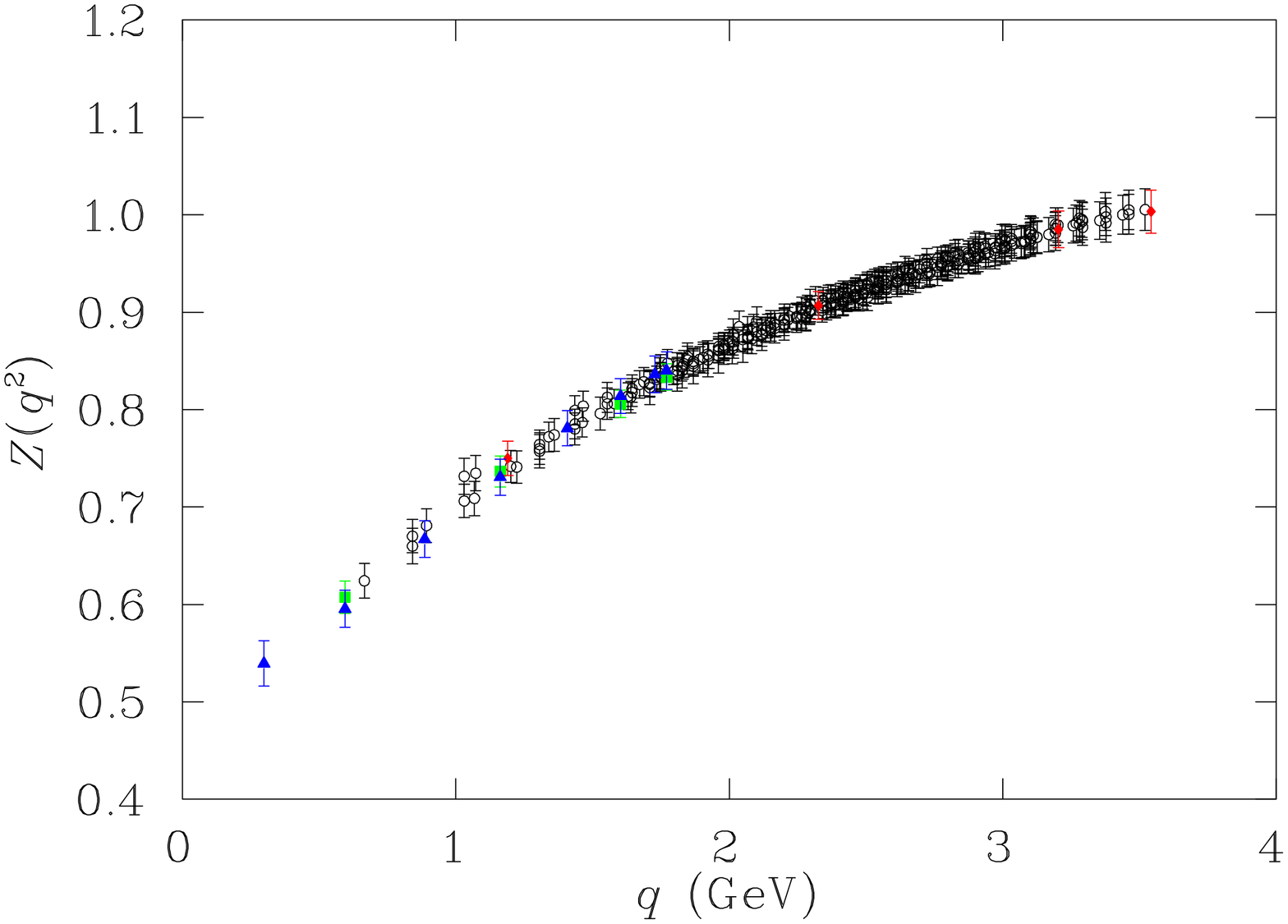,angle=90,width=8cm}
\end{center}
\caption{The quark $Z$ function with the Asqtad action in \lapIII gauge with
$ma = 0.05$.  Data
represents 76 configurations (top figure), some of which actually provide 
negative contributions.  The bottom figure shows the same data with the 
configurations producing negative contributions removed from the sample
(8 configurations).}
\label{fig:lap3_asq_Zm05}
\end{figure}

\begin{figure}[h]
\begin{center}
\epsfig{figure=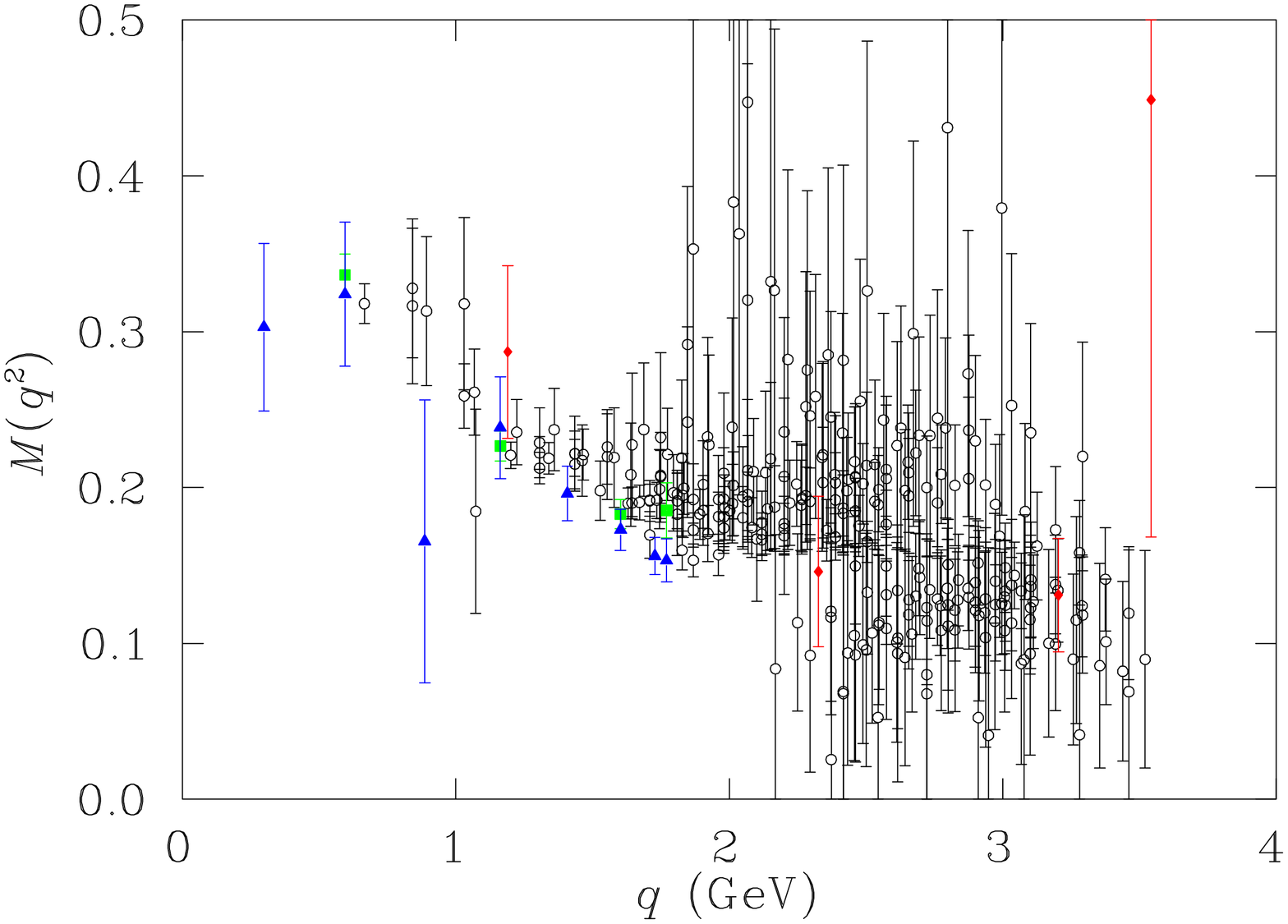,angle=90,width=8cm}
\end{center}
\begin{center}
\epsfig{figure=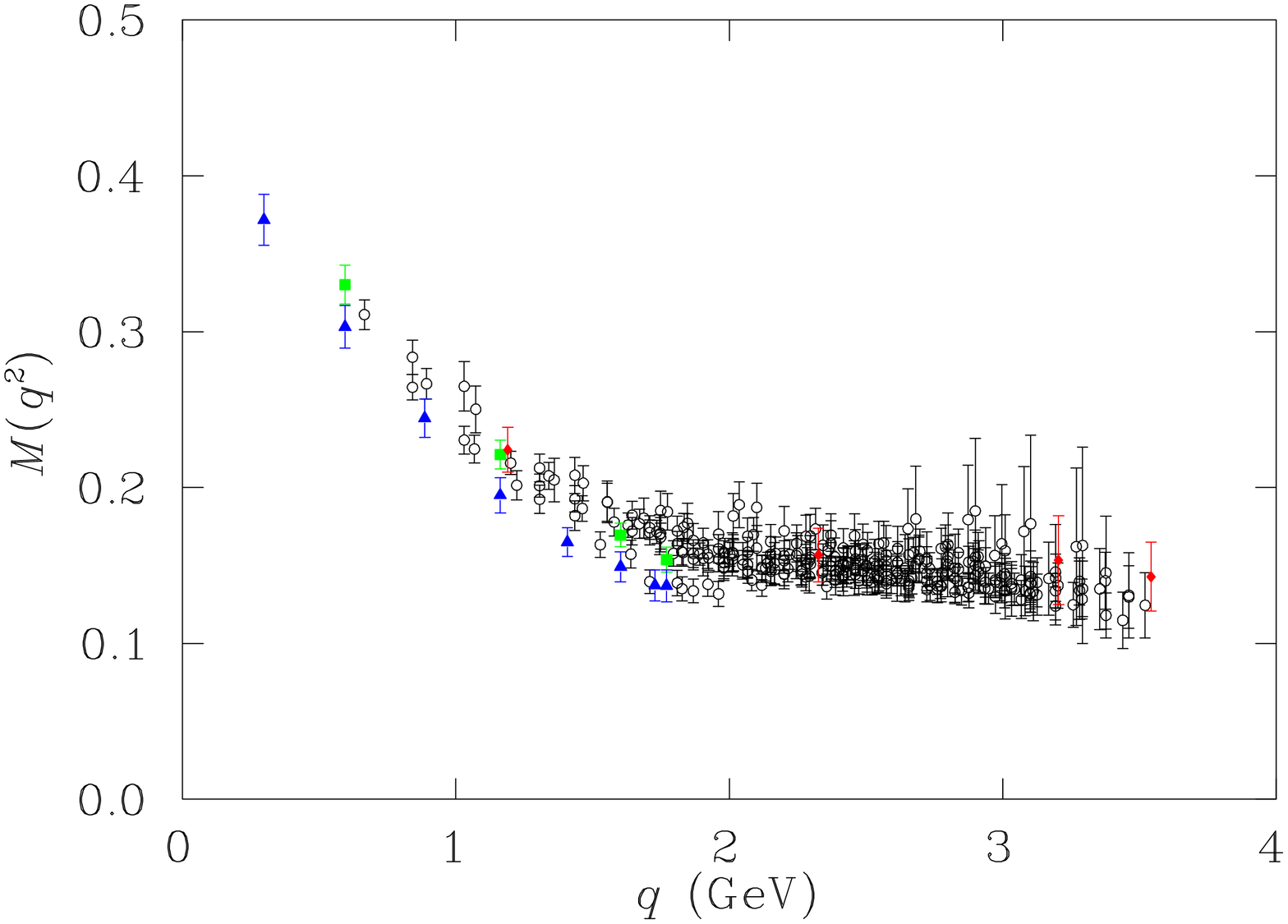,angle=90,width=8cm}
\end{center}
\caption{The quark mass function with the Asqtad action in \lapIII gauge with
$ma = 0.05$.  
The upper data represents 76 configurations, some of which actually provide 
negative contributions.  Signal is almost completely lost.
The lower data represents 68 configurations.  Removing the negative 
contributions has only barely restored the signal.}
\label{fig:lap3_asq_m05}
\end{figure}

\lapIII performs very poorly: see Figs.~\ref{fig:lap3_asq_Zm05} 
and~\ref{fig:lap3_asq_m05}.  The gauge fixing procedure failed for four of
the configurations and eight of the remaining configurations produced $Z$ and
M functions with pathological negative values.  We have seen that this type
of gauge fixing fails to produce a gluon propagator that has the correct 
asymptotic behavior~\cite{lapglu}.  It seems likely that we are dealing with 
matrices with 
vanishing determinants, which are destroying the projection onto $SU(3)$.  We
expect the degree to which this problem occurs to be dependent on the 
simulation parameters and the numerical precision used (in this work the
gauge transformations were calculated in single precision).

\section{Gauge Dependence}

\begin{figure}[h]
\begin{center}
\epsfig{figure=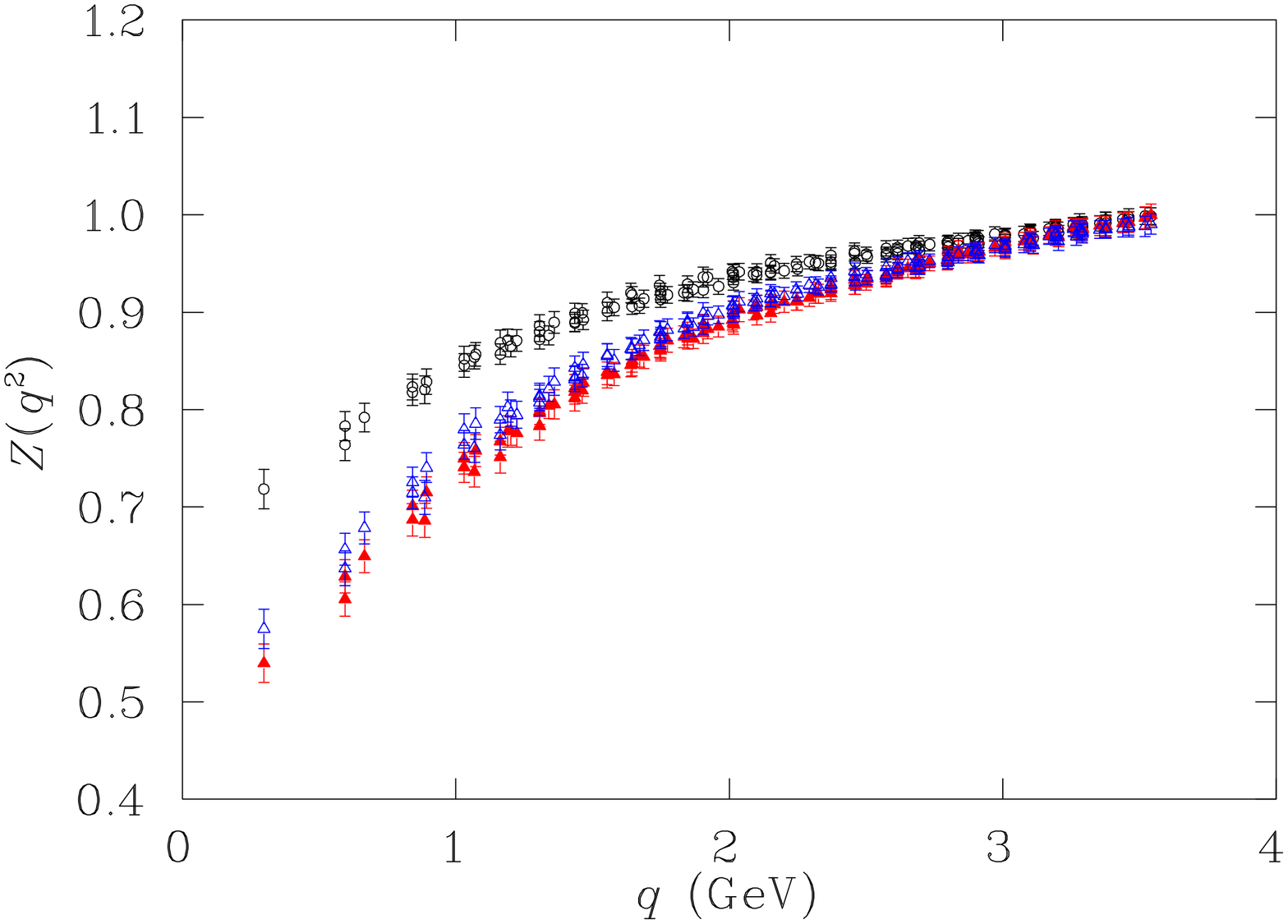,angle=90,width=8cm}
\end{center}
\caption{Comparison of the quark $Z$ functions for the Asqtad action for quark
mass $ma = 0.05$.  Points
marked with open circles are in Landau gauge, solid triangles are in 
\lapI gauge and open triangles are in \lapII gauge.  Data has been cylinder 
cut.}
\label{fig:comp_asq_Zm05}
\end{figure}

Now we investigate the quark mass and $Z$ functions in Landau, \lapI and \lapII
gauges.  Fig.~\ref{fig:comp_asq_Zm05} shows the $Z$ function for the Asqtad
action in Landau and \lapI gauges.  They are in excellent agreement in the
ultraviolet - as they ought - but differ significantly in the infrared.  The
$Z$ function in the Laplacian gauges is more strongly infrared suppressed than
in the Landau gauge.
There may be a small difference in $Z(q^2)$ between \lapI and \lapII gauges.

\begin{figure}[h]
\begin{center}
\epsfig{figure=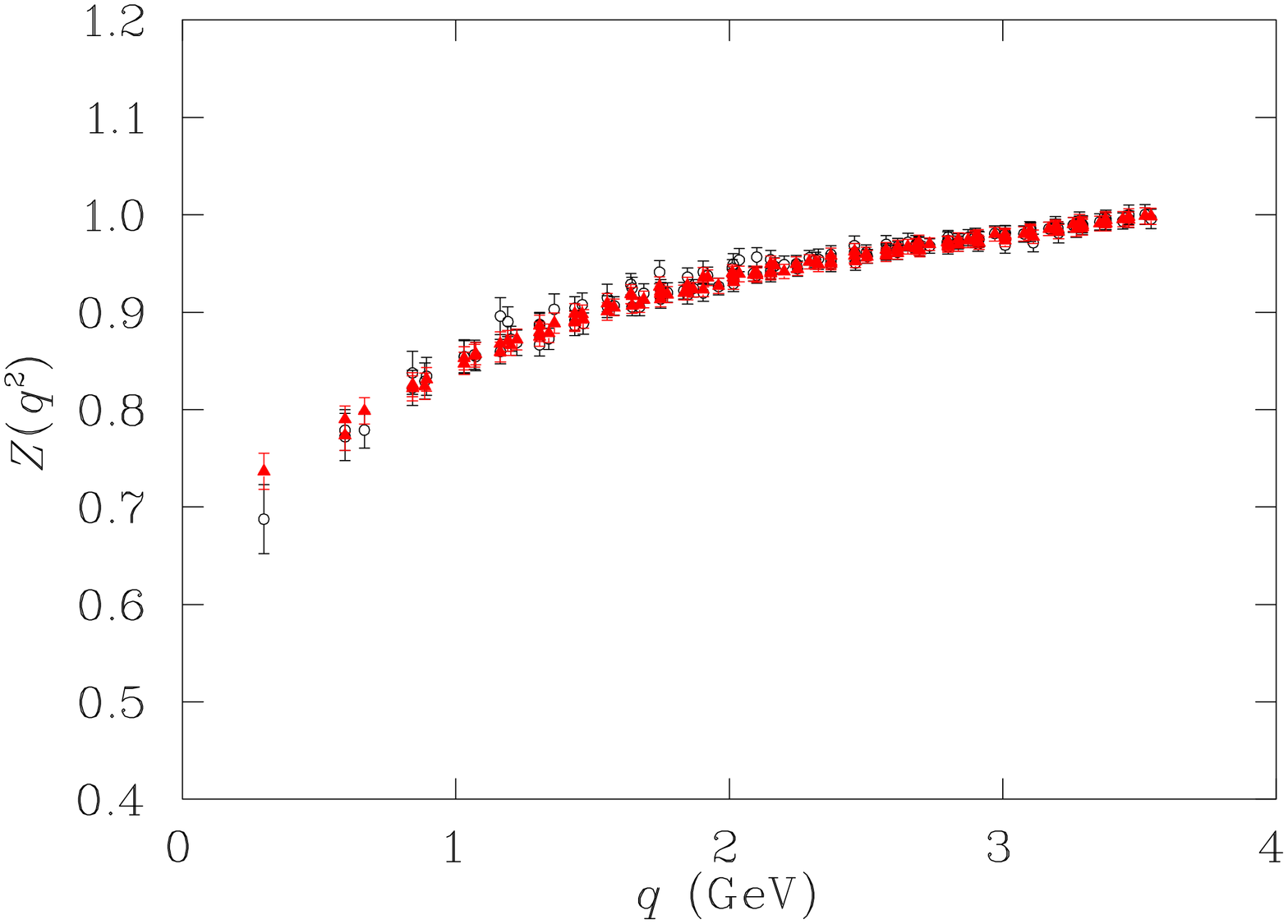,angle=90,width=8cm}
\end{center}
\caption{Comparison of the Landau gauge quark $Z$ functions for the 
heaviest ($ma = 0.075$) and the lightest ($ma = 0.0125$) quark masses,
with the Asqtad action.  Data has been cylinder cut.  We see that over this
range of values the mass dependence of the $Z$ function is very weak.}
\label{fig:lan_asq_Zcomp}
\end{figure}

\begin{figure}[h]
\begin{center}
\epsfig{figure=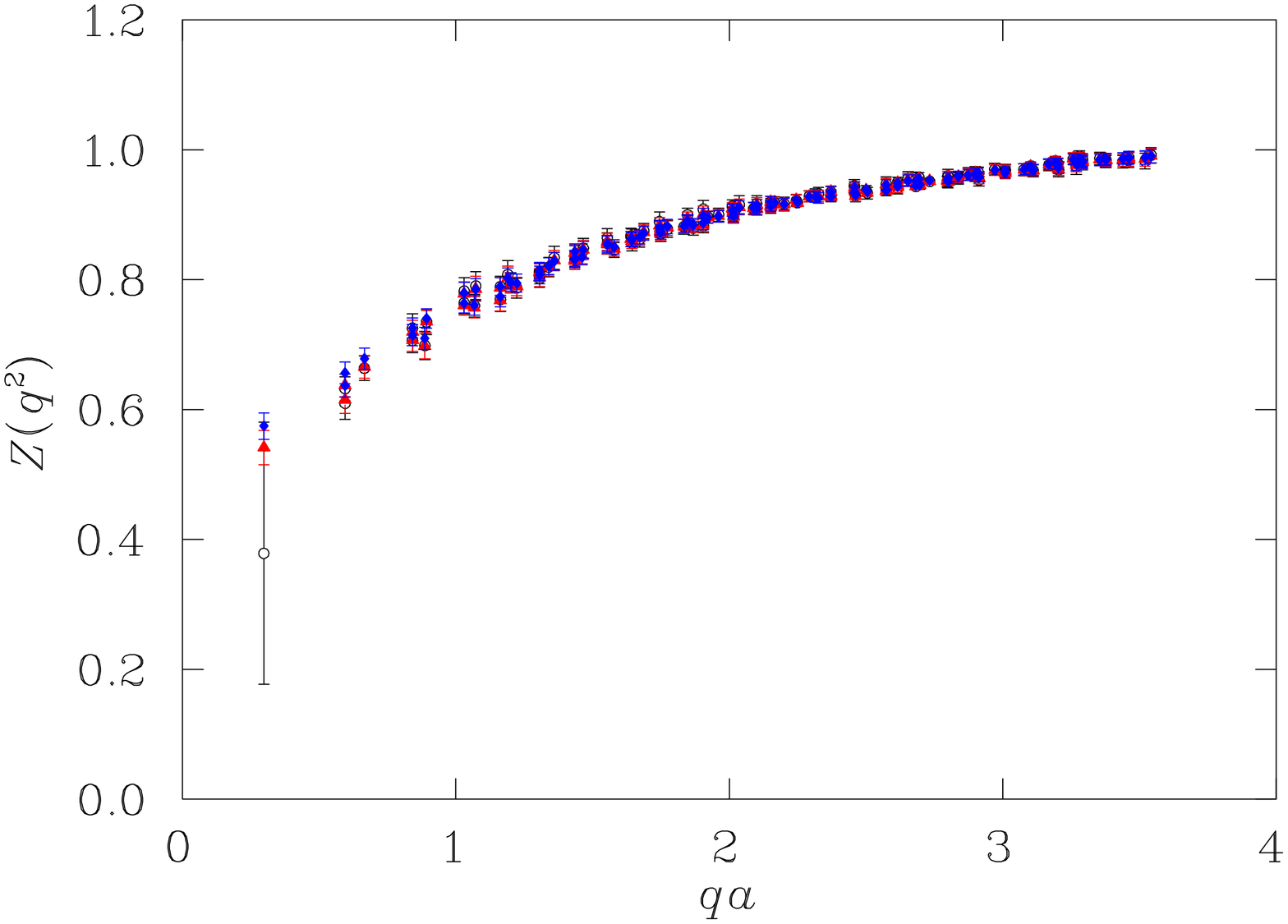,angle=90,width=8cm}
\end{center}
\caption{Comparison of the quark $Z$ functions for the three quark masses
$ma = 0.0125$, 0.025 and 0.05,
with the Asqtad action in \lapII gauge.  Data has been cylinder cut.  As in
Landau gauge, they agree to within errors, although there is a systematic
ordering of the infrared points from heaviest quark (top) to lightest 
(bottom).}
\label{fig:lap2_asq_Zcomp}
\end{figure}

In all cases the quark $Z$ function demonstrates little mass dependence.
Deviation of $Z$ from its asymptotic value of 1 is a sign of dynamical symmetry
breaking, so we expect the infrared suppression to go away in the limit of
an infinitely heavy quark.  In Fig.~\ref{fig:lan_asq_Zcomp} we show the $Z$ 
function in
Landau gauge for the lightest and the heaviest quark masses in this study.
The two are the same, to within errors, although if we look at the lowest 
momentum data we see that the point for the low mass lies below the high mass
one.  Fig.~\ref{fig:lap2_asq_Zcomp} shows $Z$ in \lapII gauge for three quark
masses.  Again, the data are consistent, to within errors, but there is a
systematic ordering of lightest to heaviest.  We conclude from this that
behavior is consistent with expectations, dependence on the quark mass 
- if any - is very weak.  One possible explanation is that all the masses 
studied are light - less than or approximately equal to the stange quark 
mass - and that heavier masses will affect the $Z$ function more clearly.

\begin{figure}[h]
\begin{center}
\epsfig{figure=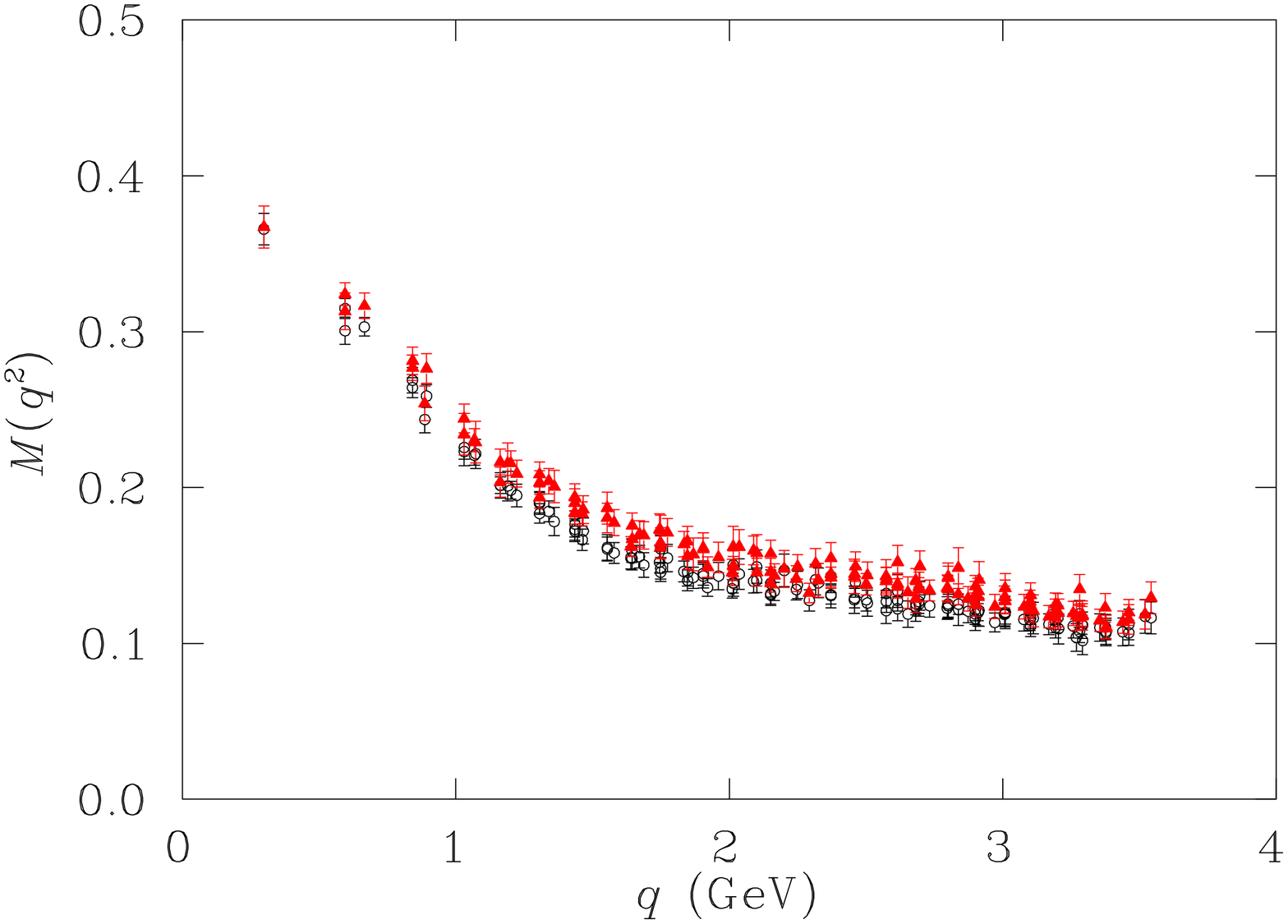,angle=90,width=8cm}
\end{center}
\caption{Comparison of the quark mass functions for the Asqtad action,
$ma = 0.05$.  Points
marked with open circles are in Landau gauge and solid triangles are in 
\lapI gauge.  Data has been cylinder cut.}
\label{fig:comp_asq_m05}
\end{figure}

\begin{figure}[h]
\begin{center}
\epsfig{figure=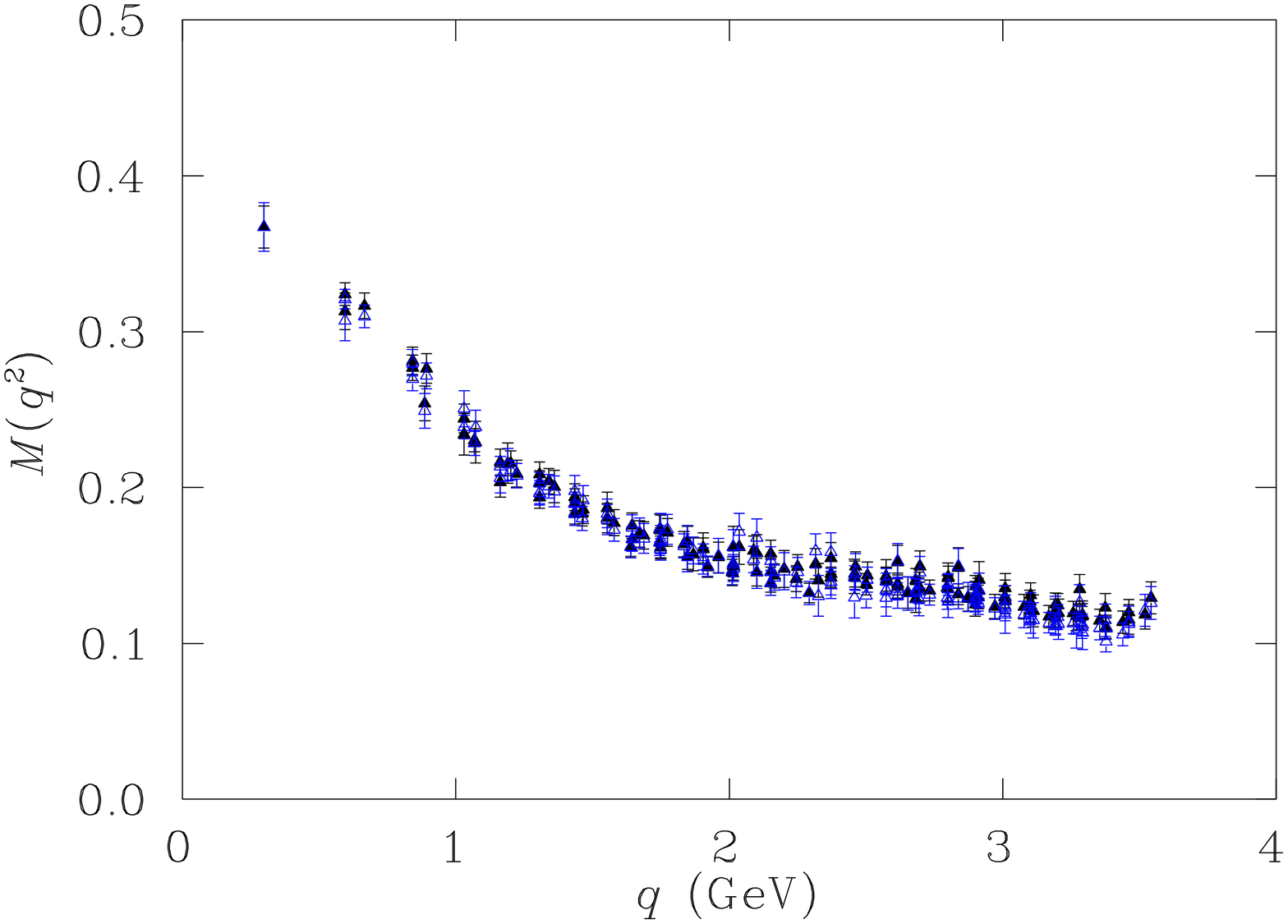,angle=90,width=8cm}
\end{center}
\caption{Comparison of the quark mass functions for the Asqtad action,
$ma = 0.05$.  Points
marked with solid triangles are in \lapI gauge and open triangles are in 
\lapII gauge.  Data has been cylinder cut.}
\label{fig:comp12_asq_m05}
\end{figure}

The mass functions in Landau and \lapI gauges - shown in 
Fig.~\ref{fig:comp_asq_m05} - agree to within errors.  The data for \lapI
gauge seems to sit a little higher than the Landau gauge through most of the
momentum range, so with greater statistics we may resolve a small difference.
The mass functions are nearly identical in \lapI 
and \lapII gauges: see Fig.~\ref{fig:comp12_asq_m05}.

\section{Modelling the Mass function}

\begin{figure}[h]
\begin{center}
\epsfig{figure=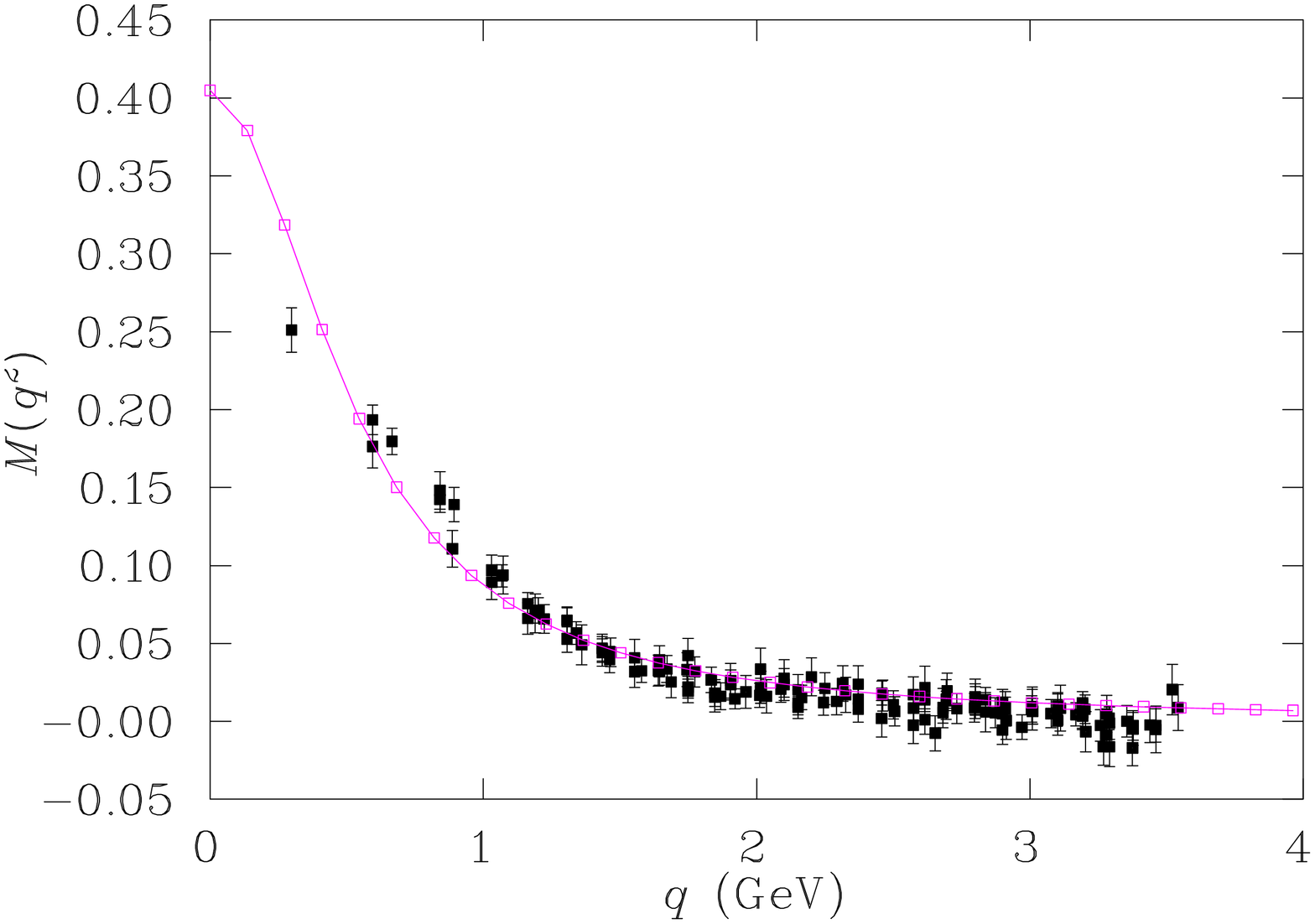,angle=90,width=7cm}
\end{center}
\begin{center}
\epsfig{figure=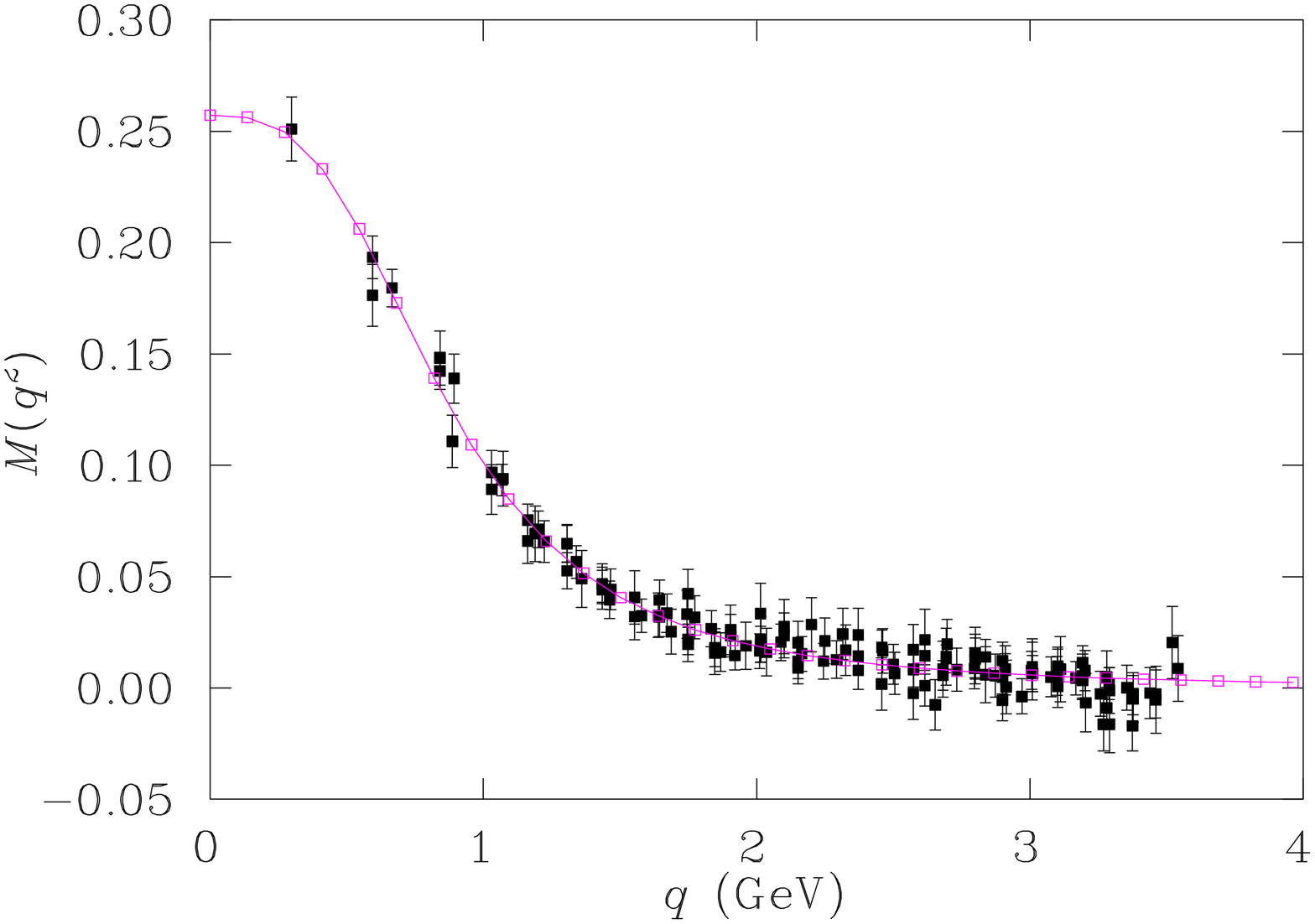,angle=90,width=7cm}
\end{center}
\caption{Mass function extrapolated to the chiral limit.  Errors are 
Jack-knife.  Fit parameters are top: c = 0.080(20), $\Lambda$ = 520(50) MeV, 
$m_0$ = 0.0, $\alpha$ = 1.0, $\chi^2$ / dof = 1.3 and
bottom: c = 0.030(4), $\Lambda$ = 870(60) MeV, $m_0$ = 0.0, 
$\alpha$ = 1.52(23), $\chi^2$ / dof = 0.49.} 
\label{fig:lan_asq_chiralfits}
\end{figure}

The Asqtad quark mass function at each value of the mass has been cylinder 
cut and
extrapolated - by a quadratic fit - to zero mass.  The quadratic fit was 
chosen on purely practical grounds and a linear fit worked almost as well.
A fit to each of the mass functions was then done, using the ansatz 
\begin{equation}
\label{eq:massfit}
M(q) = \frac{c\Lambda^{1+2\alpha}}{q^{2\alpha} + \Lambda^{2\alpha}} + m_0,
\end{equation}
which is a generalization of the one used in Ref.~\cite{Sku01a}.  As we have
seen, the quark mass function in the Laplacian gauge is almost 
indistinguishable from that in the Landau gauge, except that it is somewhat
noisier (this may change at smaller lattice spacing).  For this reason, we only
show fits in the Landau gauge.  Table~I shows a sample of the fits.  The table
is divided into two sections, one in which the parameter $\alpha$ was held 
fixed, one in which it was allowed to vary.  We see that for the heaviest 
mass, $\alpha = 1.0$ provides an excellent fit, but in the chiral limit,
$\alpha > 1.0$ is somewhat favored.  The r\^{o}le of $\alpha$ may be seen
if Fig.~\ref{fig:lan_asq_chiralfits}.  At $\alpha = 1.5$, the infrared and 
ultraviolet are significantly flatter, while the mass generation at around one 
GeV is made steeper.

\begin{table}[h]
\caption{\label{tab:lan_fits}Best-fit parameters for the ansatz, 
\eref{eq:massfit}, in Landau gauge, in physical units.  Where no errors are 
indicated, the parameter was fixed.}
\begin{ruledtabular}
\begin{tabular}{ccccccc}
\multicolumn{1}{c}{$m$} &  c  & \multicolumn{1}{c}{$\Lambda$} &
\multicolumn{1}{c}{$m_0$} & $\alpha$ & \multicolumn{1}{c}{$M(0)$} & $\chi^2$\\ 
 (MeV) &   & (MeV) & (MeV) &  & (MeV) &  per d.o.f. \\
\hline
114 & 0.35(1)  & 910(20)  & 142(7)  &    1.0   & 462(9)  & 0.38  \\
 95 & 0.36(5)  & 880(70)  & 117(7)  &    1.0   & 440(20) & 0.42  \\
 76 & 0.39(5)  & 830(70)  &  92(7)  &    1.0   & 420(20) & 0.42  \\
 57 & 0.45(4)  & 770(50)  &  70(7)  &    1.0   & 410(20) & 0.51  \\
 38 & 0.49(8)  & 720(60)  &  44(6)  &    1.0   & 400(30) & 0.56  \\
 19 & 0.54(9)  & 670(60)  &  18(6)  &    1.0   & 380(30) & 0.69  \\
  0 & 0.56(8)  & 650(50)  & -12(6)  &    1.0   & 350(20) & 0.66  \\ 
  0 & 0.80(20) & 520(50)  &  0.0    &    1.0   & 400(40) & 1.3   \\
\hline
114 & 0.28(1)  & 990(30)  & 155(7)  & 1.25(4)  & 428(7)  & 0.38  \\
 95 & 0.28(2)  & 965(40)  & 129(9)  & 1.30(10) & 404(9)  & 0.37  \\
 76 & 0.30(2)  & 930(50)  & 105(7)  & 1.29(6)  & 380(10) & 0.36  \\
 57 & 0.30(2)  & 910(40)  &  80(6)  & 1.30(2)  & 354(9)  & 0.41  \\
 38 & 0.36(4)  & 830(50)  &  54(7)  & 1.28(7)  & 350(20) & 0.46  \\
 19 & 0.30(10) & 800(200) &  29(6)  & 1.4(3)   & 310(60) & 0.55  \\
  0 & 0.30(4)  & 870(60)  &  0.0    & 1.52(23) & 260(20) & 0.49  \\  
\end{tabular}
\end{ruledtabular}
\end{table}

In this model, $\alpha$ is acting as a function of the bare mass, controlling 
the dynamical symmetry breaking.  Unfortunately, the paucity of data points in 
the infrared leaves $\alpha$ poorly determined in the chiral limit.  
Furthermore, the degree to
which our infrared data may be affected by the finite volume and by the chiral
extrapolation is not really known.  Finally, this ansatz is still crude in
that it does not provide the correct asymptotic behavior.

As was explained is Section II, the quark mass function approaches the 
renormalised quark mass in the ultraviolet, which itself becomes the bare 
mass in the $q \rightarrow \infty$ limit.  It is a general feature of this
study that the ultraviolet tail of $M(q)$ sits somewhat higher than the
bare mass.  The situation is summarised in Table~\ref{tab:massfn}.  This 
deviation from the correct asymptotic behavior is probably the consequence of
an insufficiently small lattice spacing.

\begin{table}[h]
\caption{\label{tab:massfn}
Estimates of the Landau gauge quark mass function at zero 
four-momentum.  Here $m$ is the bare input mass and $m_0$ is the ultraviolet 
mass from the fit of the mass function.}
\begin{ruledtabular}
\begin{tabular}{c|cc|cc}
    & \multicolumn{2}{c|}{$\alpha = 1.0$} 
                & \multicolumn{2}{c}{$\alpha \simeq 1.3$} \\
  $m$ (MeV) & M(0) (MeV) & $m_0 / m$ & M(0) (MeV) & $m_0 / m$\\
\hline
114 & 462(9)  & 1.25(6)  & 428(7)  & 1.35(6) \\
 95 & 440(20) & 1.23(7)  & 404(9)  & 1.36(6) \\
 76 & 420(20) & 1.21(9)  & 380(10) & 1.38(9) \\
 57 & 410(20) & 1.2(1)   & 354(9)  & 1.4(1) \\
 38 & 400(30) & 1.16(16) & 350(20) & 1.4(2) \\
 19 & 380(30) & 1.0(3)   & 310(60) & 1.5(3)\\
\end{tabular}
\end{ruledtabular}
\end{table}

\section{Conclusions}

We have seen that the Asqtad action provides a quark propagtor with improved 
rotational symmetry compared to the standard Kogut-Susskind action and that
the difference between them increases as we go to lighter quark masses.  
Furthermore, we have seen that the Asqtad action has smaller mass 
renormalization and better asymptotic behavior.

Our results for the quark propagator show that the quark mass 
function is the same, to within statistics in \lapI and \lapII gauges, while 
the $Z$ function is slightly different.  There is little difference between the
quark mass function in Landau gauge and in Laplacian gauge, but the $Z$ 
function
dips more strongly in the infrared in Laplacian gauge than in Landau gauge.
The infrared region of the Laplacian gauge 
mass function seems to be particularly badly affected by decreasing quark mass.
We have seen that the \lapIII gauge gives very poor results in $SU(3)$, in 
calculations of the quark propagator, consistent with results for the 
gluon propagator~\cite{lapglu}.  Overall the Landau gauge results of this work 
for the Asqtad appear to be
consistent within errors with the results of ealier Landau gauge studies
of the quark propagator [4,5,6].  Our results suggest that the $M$
function is insensitive to whether we use Landau or Laplacian gauge,
whereas the $Z$ function has an enhanced infrared dip in Laplacian
gauge.

As we have simulated on only one lattice, it remains to do a thorough 
examination of discretization and finite volume effects.  The chiral limit was
obtained by extrapolation, which may provide another source of systematic 
error.  It will also be interesting to investigate the \oa{4} errors by using 
an improved gluon action.  Further work may allow the development of a more 
sophisticated ansatz which has the correct asymptotic behavior.  Finally,
studies on finer lattices could be used to calculate the chiral condensate,
light quark masses and potentially $\Lambda_{\text{QCD}}$.

\begin{acknowledgments}

The authors wish to thank Derek Leinweber and Jonivar Skullerud for useful
discussions.  The work of UMH and POB was supported in part by DOE contract 
DE-FG02-97ER41022. The work of AGW was supported by the Australian Research 
Council.

\end{acknowledgments}

\appendix*
\section{Staggered quark propagators}

In this appendix we give details of the quark propagator calculation using
the Kogut-Susskind and Asqtad actions.
The free KS action is
\begin{multline}
S = \frac{1}{2} \sum_{x,\mu} \chibar(x) \eta_\mu(x) 
   \bigl( \chi(x+\mu) - \chi(x-\mu) \bigr) \\ 
   + m\sum_x \chibar(x)\chi(x)
\end{multline}
where the staggered phases are: $\eta_\mu(x) = (-1)^{\zeta^{(\mu)} \cdot x}$ 
and
\begin{equation}
\zeta^{(\mu)}_\nu = \biggl\{ \begin{array}{ll}
			1 & \mbox{if $\nu < \mu$} \\
			0 & \mbox{otherwise.}
		 \end{array}
\end{equation}
To Fourier transform, write 
\begin{equation}
k_\mu = \frac{2\pi n_\mu}{L_\mu} \quad | \quad n_\mu = 0,\ldots,L_\mu-1
\end{equation}
as $k_\mu = p_\mu + \pi\alpha_\mu$, where
\begin{gather}
\label{eq:stag_mom}
p_\mu = \frac{2\pi m_\mu}{L_\mu} \quad | 
   \quad m_\mu = 0,\ldots,\frac{L_\mu}{2}-1 \\
\alpha_\mu = 0,1,
\end{gather}
and define $\int_k \equiv \frac{1}{V} \sum_k$.  Then
\begin{gather}
\int_k = \int_p \sum_{\alpha_\mu = 0}^1 \\
\chi(x) = \int_k e^{ik \cdot x}\chi(k) = \int_p \sum_{\alpha} 
   e^{i(p+\pi\alpha) \cdot x} \chi_\alpha(p).
\end{gather}
Defining
\begin{align}
\deltabar_{\alpha\beta} & = \Pi_\mu \delta_{\alpha_\mu\beta_\mu |\bmod 2} \\
{(\gambar_\mu)}_{\alpha\beta} & = (-1)^{\alpha_\mu} 
	\deltabar_{\alpha+\zeta^{(\mu)},\beta}
\end{align}
where the $\gambar_\mu$ satisfy
\begin{gather}
\{\gambar_\mu, \gambar_\nu\} = 2 \delta_{\mu\nu} \deltabar_{\alpha\beta} \\
\gambar_\mu^\adj = \gambar_\mu^T = \gambar_\mu^* = \gambar_\mu,
\end{gather}
forming a ``staggered'' Dirac algebra.
Putting all this together, we can derive a momentum space expression for the KS
action,
\begin{equation}
S = \int_p \sum_{\alpha\beta} \chibar_\alpha(p)
    \Bigl[ i \sum_\mu {(\gambar_\mu)}_{\alpha\beta} \sin(p_\mu)
    + m\deltabar_{\alpha\beta} \Bigr] \chi_\beta(p).
\end{equation}

From this we can see that, in momentum space, the tadpole improved, tree-level 
form of the quark propagator is
\begin{equation}
S_{\alpha\beta}^{-1}(p;m) = u_0 i \sum_\mu {(\gambar_\mu)}_{\alpha\beta} 
   \sin(p_\mu) + m\deltabar_{\alpha\beta}
\end{equation}
where $p_\mu$ is the discrete lattice momentum given by 
Eq.~\eqref{eq:stag_mom}.  
Assuming that the full propagator retains this form (in analogy to the 
continuum case) we write
\begin{align}
S_{\alpha\beta}^{-1}(p) &= i\sum_\mu {(\gambar_\mu)}_{\alpha\beta} 
   \sin(p_\mu) A(p) + B(p) \deltabar_{\alpha\beta} \\
	&= Z^{-1}(p) \Bigl[i\sum_\mu {(\gambar_\mu)}_{\alpha\beta} \sin(p_\mu) 
		+ M(p) \deltabar_{\alpha\beta} \Bigr].
\end{align}
For the KS action, it is convenient to define
\begin{equation}
q_\mu \equiv \sin(p_\mu)
\end{equation}
as a shorthand.

Numerically, we calculate the quark propagator in coordinate space,
\begin{align}
G(x,y) 	&= \bigl\langle \chi(x)\chibar(y) \bigr\rangle \notag\\
       	&= \sum_{\alpha\beta} \int_{p,r} \exp\bigl\{ i(p+\pi\alpha)x 
	- i(r+\pi\beta)y \bigr\} \notag\\ 
	  &\qquad\qquad \times \bigl\langle\chi_\alpha(p) \chibar_\beta(r) 
	  \bigr\rangle \notag\\
       	&= \sum_{\alpha\beta} \int_{p,r} \exp\bigl\{ i(p+\pi\alpha)x 
	- i(r+\pi\beta)y \bigr\}  \notag\\
	  &\qquad\qquad \times \delta_{pr} S_{\alpha\beta}(p) \notag\\
       	&= \sum_{\alpha\beta} \int_p \exp\bigl\{ ip(x-y) \bigr\} \notag\\
	  &\qquad\qquad \times \exp\bigl\{ i\pi(\alpha x - \beta y) \bigr\} 
	  S_{\alpha\beta}(p).
\end{align}
To obtain the quark propagator in momentum space, we take the Fourier transform
of $G(x,0)$ and, decomposing the momenta
as $k_\mu = r_\mu + \pi\delta_\mu \quad | \quad 0 \le r_\mu < \pi$, get
\begin{align}
G(k) & = G(r+\pi\delta) \equiv G_\delta(r) = \sum_x e^{-ikx} G(x,0) \notag\\
	&= \sum_{\alpha\beta} \int_p \sum_x \exp\bigl\{ -i(r+\pi\delta)x 
	\bigr\} \notag\\
	  &\qquad\qquad \times \exp\bigl\{ i(p+\pi\alpha)x \bigr\} 
	  S_{\alpha\beta}(p) \notag\\
	&= \sum_{\alpha\beta} \int_p \delta_{pr}\deltabar_{\alpha\delta}
	S_{\alpha\beta}(p).
\end{align}
Thus, in terms of the KS momenta, $q$, 
\begin{equation}
G_\delta(q) = \sum_\beta S_{\delta\beta}(q) 
	    = Z(q) \frac{-i\sum_\mu (-1)^{\delta_\mu} q_\mu + M(q)}
		{q^2 + M^2(q)},
\end{equation}
from which we obtain
\begin{align}
\sum_\delta \tr G_\delta(q)  
	&= 16N_c \frac{Z(q)M(q)}{q^2 + M^2(q)} \notag\\
	&= 16N_c {\cal B}(q),
\end{align}
and
\begin{align}
i\sum_\delta\sum_\rho (-1)^{\delta_\rho} q_\rho \tr [G_\delta(q)]
	&= 16N_c q^2 \frac{Z(q)}{q^2 + M^2(q)} \notag\\
	&= 16N_c q^2 {\cal A}(q).
\end{align}

Note: we could determine
\begin{align}
A(p) &= Z^{-1}(p) \notag\\ 
     &= \frac{-i}{16N_c \sum_\nu \sin^2(p_\nu)} \notag\\
	&\quad\quad \times \sum_{\alpha\beta} \sum_{\rho} 
	{(\gambar_\rho)}_{\alpha\beta} \sin(p_\rho) 
	\tr \bigl[ S^{-1}_{\beta\alpha}(p) \bigr] \\
B(p) &= \frac{M(p)}{Z(p)} = \frac{1}{16N_c} \sum_\alpha 
	\tr \bigl[ S^{-1}_{\alpha\alpha}(p) \bigr],
\end{align}
but we would rather avoid inverting $S_{\alpha\beta}(p)$.

Putting it all together we get
\begin{align}
A(q) &= Z^{-1}(q) = \frac{{\cal A}(q)}{{\cal A}^2(q) q^2
	+ {\cal B}^2(q_\mu)} \\
B(q) &= \frac{M(q)}{Z(q)} = \frac{{\cal B}(p)}
	{{\cal A}^2(q) q^2 + {\cal B}^2(p)} \\
M(q) &= \frac{{\cal B}(q)}{{\cal A}(q)}.
\end{align}

The tadpole improved, tree-level behavior of the $Z$ and mass functions are
simply
\begin{equation}
Z^0 = \frac{1}{u_0}
\end{equation} 
and
\begin{equation}
M^0 = \frac{m}{u_0}
\end{equation}
respectively.

The Asqtad quark action~\cite{Org99} is a fat-link Staggered action using
three-link, five-link and seven-link staples along with the three-link Naik
term~\cite{Nai89} and five-link Lepage term~\cite{Lep99}, with tadpole 
improved coefficients tuned to remove all tree-level \oa{2} errors.  This 
action was motivated by the desire to minimize quark flavour changing 
interactions, but has also been reported to have good rotational symmetry.

At tree-level (i.e. no interations, links set to the identity), the staples
in this action make no contribution, so the action reduces to the Naik action.
\begin{multline}
S = \frac{1}{2} \sum_{x,\mu} \chibar(x) \eta_\mu(x) \Bigl[
   \frac{9}{8} \bigl( \chi(x+\mu) - \chi(x-\mu) \bigr) \\ - 
   \frac{1}{24} \bigl( \chi(x+3\mu) - \chi(x-3\mu) \bigr) \Bigr]
   + m\sum_x \chibar(x)\chi(x)
\end{multline}
The quark propagator with this action has the tree-level form
\begin{multline}
S_{\alpha\beta}^{-1}(p;m) = u_0 i \sum_\mu {(\gambar_\mu)}_{\alpha\beta}
     \sin(p_\mu) \Bigl[\frac{9}{8} \sin(p_\mu) \\
       - \frac{1}{24} \sin(3p_\mu) \Bigr] 
     + m\deltabar_{\alpha\beta}
\end{multline}
so we choose
\begin{align}
q_\mu(p_\mu) &\equiv \frac{9}{8} \sin(p_\mu) - \frac{1}{24} \sin(3p_\mu) \\ 
             &= \sin(p_\mu) \Bigl[ 1 + \frac{1}{6} \sin^2(p_\mu) \Bigr].
\end{align}
Having identified the correct momentum for this action, we can calculate the
invariant functions as before.  No further tree-level correction is required.


\end{document}